\documentclass[12pt]{article}
\usepackage{amsfonts,amsmath}
\usepackage{amssymb}
\usepackage{mathrsfs}
\usepackage[hypertex] {hyperref}

\newcommand{\dr}{{{\rm d}}}

\newcommand{\ord}{\widehat{\mathsf{O}}}
\renewcommand{\theequation}{\thesection.\arabic{equation}}

\makeatletter \@addtoreset{equation}{section} \makeatother

{\vspace{3mm} }

\def\al{\alpha}

\def\*{\star}

\def\E2{\mathbf{E}}

\newcommand{\be}{\begin{equation}}
\newcommand{\ee}{\end{equation}}
\newcommand{\bee}{\begin{eqnarray}}
\newcommand{\beee}{\begin{array}}
\newcommand{\eee}{\end{eqnarray}}
\newcommand{\eeee}{\end{array}}
\newcommand{\gb}{\beta}

\newcommand{\gd}{\delta}

\newcommand{\gs}{\sigma}
\newcommand{\go}{\omega}

\newcommand{\p}{\partial}

\newcommand{\ff}{\frac}

\begin{document}
\begin{flushright}
FIAN/TD/19-2022\\
\end{flushright}

\vspace{0.5cm}
\begin{center}
{\large\bf On $z$-dominance, shift symmetry and spin locality in
higher-spin theory}

\vspace{1 cm}

\textbf{V.E.~Didenko and A.V.~Korybut}\\

\vspace{1 cm}

\textbf{}\textbf{}\\
 \vspace{0.5cm}
 \textit{I.E. Tamm Department of Theoretical Physics,
Lebedev Physical Institute,}\\
 \textit{ Leninsky prospect 53, 119991, Moscow, Russia }\\

\par\end{center}

\begin{center}
\vspace{0.6cm}
% didenko@lpi.ru, vasiliev@lpi.ru \\

\par\end{center}

\vspace{0.4cm}

\begin{abstract}
\noindent The paper aims at the qualitative criterion of
higher-spin locality. Perturbative analysis of the Vasiliev
equations gives rise to the so-called $z$-dominated non-localities
which nevertheless disappear from interaction vertices leaving the
final result spin-local in all known cases. This has led one to
the $z$ -- dominance conjecture that suggests universality of the
observed cancellations. Here we specify conditions which include
observation of the higher-spin shift symmetry and prove validity
of this recently proposed conjecture. We also define a class of
spin-local and shift-symmetric field redefinitions which is argued
to be the admissible one with respect to spin-locality.
\end{abstract}

\newpage
\tableofcontents
\newpage
\section{Introduction}
Higher spin (HS) gauge theory describes interactions between
massless fields of all spins. This theory is not local in the
common sense since the higher the spin of the fields involved in a
vertex the higher the number of space-time derivatives entering
this vertex
\cite{Bengtsson:1983pd,Berends:1984wp,Fradkin:1987ks,Fradkin:1991iy}.
From this perspective the conventional notion of locality is not
an option, however one may wonder if there can be finitely many
derivatives in HS interaction vertex that contains fields of fixed
spins. In \cite{Gelfond:2018vmi} the notion of {\it spin locality}
in four dimensions was introduced. The vertex is said to be spin
local if it contains finite number of contractions between
undotted or dotted spinorial indices of the zero-form module $C$
(to be clarified below). In the lowest orders of perturbation
theory the spin locality can be translated to a finite amount of
space-time derivatives in the vertex\footnote{For higher orders
the relation gets more complex. However, in the recent paper
\cite{Vasiliev:2022med} an additional requirement for spin local
vertices called the {\it projective spin locality} was introduced.
It allows one to state that the respective vertex contains
finitely many space-time derivatives. Discussion of this aspect is
beyond the scope of this paper.}. In the recent paper
\cite{Didenko:2020bxd} the obtained $C^3$ contributions were
claimed to lead to the spin-local holomorphic vertices by virtue
of the so called $z$ -- dominance conjecture
\cite{Gelfond:2018vmi}\footnote{Called a lemma in that reference
since it was proven at order $C^2$.} (see also
\cite{Gelfond:2019tac}, \cite{Didenko:2019xzz}). One of these
vertices was found then in a manifestly spin-local form
\cite{Gelfond:2021two} providing evidence for the claim made in
\cite{Didenko:2020bxd}. The main goal of this paper is to give the
rigorous proof for this claim by specifying conditions of the $z$
-- dominance conjecture and extend it to higher orders. Before
formulating the problem in detail we provide a brief overview on
HS theory at its current stage covering only those aspects which
are necessary to this paper. For more detailed reviews on higher
spins we address the reader to \cite{Vasiliev:1999ba},
\cite{Bekaert:2004qos}, \cite{Didenko:2014dwa},
\cite{Ponomarev:2022vjb}.

So far the full nonlinear dynamics of higher spin fields is
available at the level of equations of motion
\cite{Vasiliev:1992av} only. These equations are the generating
ones that allow one order by order reconstructing dynamics of HS
fields in the following, so-called, unfolded
\cite{Vasiliev:1988xc}, \cite{Vasiliev:1988sa} form
\begin{equation}\label{1form}
\dr_x \go +\go\ast
\go=\Upsilon(\go,\go,C)+\Upsilon(\go,\go,C,C)+\ldots ,
\end{equation}
\begin{equation}\label{0form}
\dr_x C+\omega \ast C-C\ast
\omega=\Upsilon(\go,C,C)+\Upsilon(\go,C,C,C)+\ldots.
\end{equation}
Here $\omega$ is a HS potential one-form, while $C$ is a zero-form
containing matter fields along with HS Weyl tensors and their
on-shell derivatives. Both functions are formal power series in
the two-component spinorial variables $y$ and $\bar{y}$
\begin{equation}
\omega(Y\vert x)=\sum_{m,n} \dr x^\mu \omega_{\mu\, \alpha_1\ldots
\alpha_m,\dot{\alpha}_1\ldots \dot{\alpha}_n}y^{\alpha_1}\ldots
y^{\alpha_m} \bar{y}^{\dot{\alpha}_1}\ldots
\bar{y}^{\dot{\alpha}_n}\,,
\end{equation}
\begin{equation}
C(Y\vert x)=\sum_{m,n}  C_{\alpha_1\ldots
\alpha_m,\dot{\alpha}_1\ldots \dot{\alpha}_n}y^{\alpha_1}\ldots
y^{\alpha_m} \bar{y}^{\dot{\alpha}_1}\ldots
\bar{y}^{\dot{\alpha}_n}\,.
\end{equation}
Spin $s$ field and all its derivatives are encoded in $\dr
x^{\mu}\omega_{\mu\, \alpha_1\ldots \alpha_m,\dot{\alpha}_1\ldots
\dot{\alpha}_n}$ for $m+n=2(s-1)$ and in $C_{\alpha_1\ldots
\alpha_m,\dot{\alpha}_1\ldots \dot{\alpha}_n}$ for $|m-n|=2s$.
Functions $\omega$ and $C$ are subject to star product defined as
\begin{equation}
f(y,\bar{y})\ast g(y,\bar{y})=\int \frac{\dr^2 u\, \dr^2 v\, \dr^2
\bar{u}\, \dr^2 \bar{v}}{(2\pi)^4}\, e^{iu_\alpha v^\alpha
+i\bar{u}_{\dot{\alpha}}\bar{v}^{\dot{\alpha}}}
f(y+u,\bar{y}+\bar{u})g(y+v,\bar{y}+\bar{v})\,.
\end{equation}
In what follows the star product with respect to $\bar y$ variable
\be\label{barstar}
f(y,\bar{y})\bar \ast g(y,\bar{y})=\int \frac{\dr^2 \bar{u}\,
\dr^2 \bar{v}}{(2\pi)^2}\,
e^{i\bar{u}_{\dot{\alpha}}\bar{v}^{\dot{\alpha}}}
f(y,\bar{y}+\bar{u})g(y,\bar{y}+\bar{v})
\ee
is persistently present in all products of fields $\go$ and $C$'s,
so we will omit symbol $\bar{*}$ for brevity.

Vasiliev system has an enlarged amount of auxiliary spinor
variables. In addition to $y,\bar{y}$ it contains functions that
depend also on $Z=(z,\bar{z})$. However, field equations are
designed in such a way that $z$-dependence vanishes upon plugging
master-fields into those equations that govern HS dynamics and
result in \eqref{1form}-\eqref{0form}. In addition, the system has
a closed holomorphic sector for which the field dependence on
$\bar z$ is not relevant. We stay in that sector in which
\eqref{barstar} remains undeformed by the holomorphic HS
interactions, while $\bar z$ never shows up.

\subsection{Problem setting and the results}
HS vertices can be systematically extracted from the Vasiliev
generating equations by solving for master fields $W(z,y|x)$ and
$B(z,y|x)$ order by order \cite{Vasiliev:1992av},
\cite{Sezgin:2002ru} (see also \cite{Didenko:2015cwv}). The
vertices from the right hand sides of \eqref{1form} and
\eqref{0form} then appear as
\begin{align}
&\Upsilon(\go, \go, C,\dots, C):=-\dr_x W-W*W\,,\label{ver1}\\
&\Upsilon(\go, C,\dots, C):=-\dr_x B-[W,B]_*\,.\label{ver0}
\end{align}
Note that while $W$ and $B$ do depend manifestly on $z$, the
resulting vertices $\Upsilon$ are $z$ -- independent by the
Vasiliev construction. Solving Vasiliev generating system in a
spin-local way at least up to $O(C^3)$ one finds
\cite{Didenko:2020bxd},\cite{Didenko:2019xzz},\cite{Didenko:2018fgx} that vertices reduce to a sum of $z$ --
dependent expressions that look as follows\footnote{From now
onwards we omit star product $\bar{*}$ which is invisibly present
in between fields $\go$ and $C$'s.}
\begin{equation}\label{contr}
\int_0^1\dr\mathcal{T} \int \dr\rho_1\ldots \dr\rho_N \,
\pi(\partial_\go,\partial_j,y,z|\,\rho_i,\mathcal{T})\omega(\mathcal{T}\rho_1
z+\ldots) C(\mathcal{T}\rho_2 z+\ldots)\ldots C(\mathcal{T}\rho_N
z+\ldots) e^{i\mathcal{T} z_\alpha y^\alpha}.
\end{equation}
Here $\pi(\partial_\go,\partial_j,y,z|\,\rho_i,\mathcal{T})$ is
polynomial in spinorial arguments
$\{\partial_\omega,\partial_j,y,z\}$ with $\partial_\go$ and
$\partial_j$ being the derivatives with respect to the full
holomorphic spinorial argument of field $\omega$ and the $j$-th
field $C$ as seen from left to right. The integration over
$\rho$'s is carried over a compact domain. Note that $\go$ may
appear at any positions among $C$'s, while in \eqref{contr} it is
on the left of $C$'s as a matter of example. Dots in arguments of
$\go$ and $C$ contain no $z$'s. These may include however various
$\p_i$. In this case the corresponding contribution can be
apparently non-local. Note that while the $z$ -- dependence is
clearly present in each contribution \eqref{contr}, the sum of all
of these should be $z$ -- independent. Now we are getting to the
proposal of the mechanism that annihilates $z$ and makes vertex
local. In the sequel we will use the following shorthand notation
of \cite{Didenko:2020bxd,Didenko:2019xzz,Didenko:2018fgx}
\be
t_\alpha=-i\partial_\omega{}_\alpha\,,\qquad
p_{j\alpha}=-i\partial_{j\alpha}\,.
\ee

\paragraph{$Z$ -- dominance conjecture}

Formulated  for the first time in \cite{Gelfond:2018vmi} and
dubbed the $z$ -- dominance lemma (see footnote 2), it rests on
the following observation. The $z$-dependence of HS vertex (a sum
of expressions \eqref{contr}) is trivial provided it acquires a
form of a total derivative with respect to $\mathcal{T}$ vanishing
at $\mathcal{T}=1$. Specifically, the integral boils down to its
$\mathcal{T}$-boundary which consists of two points
$\mathcal{T}=0$ and $\mathcal{T}=1$. The final result is then $z$
-- independent if the non-zero contribution comes from
$\mathcal{T}=0$ only. Now, if the integrand \eqref{contr} depends
on some parameter $\xi$ via $\exp(iF(\mathcal{T},\rho_1,\ldots
\rho_N)\xi)$\footnote{Such exponential dependence naturally
results from \eqref{contr} by rewriting $\go$ and $C$ field
arguments using translation operator, e.g.
$C(y)=e^{iy^{\al}p_{\al}'}C(y)\Big|_{y'=0}$.}, where
\begin{equation}\label{Ex}
0\leq F(\mathcal{T},\rho_1,\ldots \rho_N)\leq \mathcal{T},
\end{equation}
then the result of the integration followed by summation of all
contributions should not depend on parameter $\xi$. The latter
assertion is the essence of the $z$ -- dominance conjecture of
\cite{Gelfond:2018vmi},\cite{Gelfond:2019tac}. Terms satisfying
\eqref{Ex} are called the {\it $z$-dominated}. {Particularly,
$\xi$ may contain contractions $\xi=p_ip_j$, while the statement
implies that even though $pp$-contractions show up in apparently
non-local way, the resulting vertex do not depend on them at all.
} At order $C^2$ this statement was explicitly proven in
\cite{Didenko:2018fgx} and conjectured to hold at higher orders.
The rationale behind it is exponential $\exp(i\mathcal{T}z_\alpha
y^\alpha)$ of \eqref{contr}, which seemingly suggests that the
result of integration over $\mathcal{T}$ and $\rho_i$ contains all
powers of $zy$, whereas no power of $zy$ can contribute to the
vertex since it is $z$-independent by definition. It is then
natural to expect that exponential $\exp(i F\cdot\xi)$ if present
in the integrand can not contribute to the $z$-independent
expression either as soon as it is dominated by \eqref{Ex}.

The $z$ -- dominance conjecture plays an important role in the HS
locality problem. In particular, the results of
\cite{Didenko:2020bxd}, where spin-locality was analyzed up to
order -- $C^3$ heavily rely on it as the main idea was to solve
the Vasiliev equations in such a way that the non-local
contractions appeared in a {$z$-dominated} form. However, the
conjecture as formulated in \cite{Gelfond:2018vmi} seems to be too
strong beyond $C^2$ and should be replaced with a milder
statement. Namely, parameters that enter the integrand in a
$z$-dominated form can contribute to the final result no more than
polynomially. The latter formulation will be regarded here as the
{\it $z$ -- dominance conjecture}. The goal of this paper is to
specify conditions that make this conjecture valid.

Even though the content of the $z$ -- dominance conjecture looks
clear there is a tricky point here! In bringing vertex to a
manifestly $z$-independent form one has to use the Schouten
identities and partial integration over $\rho_i$. The latter
brings various $p_j$ to the pre-exponential (to be clarified
below) upon partial integration. This process can go on
indefinitely, terminating in finite or infinite steps before $z$
eventually vanishes. When total amount of such steps is finite
then the conjecture indeed implies spin locality but in general it
might take infinite steps. In the latter case the originally
harmless and finite in spinorial derivatives pre-exponential
$\pi(\partial_\go,\partial_j,y,z|\,\rho_i,\mathcal{T})$ may turn
into a power series in general. For example in a recent paper
\cite{Gelfond:2021two} there was no evidence that this process
stops no sooner than the highest power of various spinorial
derivatives entering the pre-exponential reaches six. As one can
expect some sort of a symmetry should be responsible for such a
truncation.

A certain shift symmetry manifesting itself at least up to order
$C^3$ is indeed available within the Vasiliev approach\footnote{In
\cite{Didenko:2022qga} the shift symmetry was proven to all
orders. However, the generating equations used there differed from
the original Vasiliev ones.} as we show. The symmetry is realized
as shifts of various derivatives over spinorial holomorphic
arguments entering a given vertex by a constant $\p\to\p+const$.
Its action results in a remarkably simple transformation of HS
vertices. These shifts are of the type present in the so-called
structure lemma introduced in \cite{Gelfond:2018vmi} and playing
an important role in the HS locality theorems. The effects of the
shifts were analyzed from the locality stand point in
\cite{Gelfond:2018vmi}, i.e. up to the effects caused by
$\p_{\go}$ -- shift of field $\omega$ which could not impact
locality. However, taking $\p_{\go}$ -- shift into account turns
out to be very useful giving rise to the exact shift symmetry of
HS vertex as we observe. The symmetry will play an important role
in the analysis carried out in this paper.

The main result of our investigation can be formulated as follows:
\vspace{3mm}

{\it If non-local contractions entering various Lorentz invariant
and shift-symmetric contributions to a HS vertex are $z$-dominated
then the corresponding vertex is spin-local.}\vspace{3mm}

In proving the $z$ -- dominance conjecture we were able to relax
neither the shift-symmetry nor the Lorentz invariance requirements
as they both appear to play an important role in our proof of this
statement. As an application of our analysis we provide an
argument that a certain class of local and shift-symmetric field
redefinitions defined in our paper should respect locality of HS
vertices.

\subsection{Sketching the main idea}
There are few steps and observations behind the proof of the $z$
-- dominance conjecture that we would like to bring to reader's
notice before proceeding to technical details.

Following \cite{Didenko:2020bxd} we start by noting  that in order
to extract HS vertices from the Vasiliev equations it is
sufficient to solve them up to the $z$-dominated terms. A
clarification here is required. Dropping these terms off clearly
ruins consistency as the resulting vertex becomes no longer
$z$-independent. However, consistency along with the
$z$-independence can be restored by acting on the 'reduced vertex'
with a special operator called $\ord_{-\infty}$ (to be explained).
Although this step looks purely technical, dropping the dominated
terms leads to a substantial simplification in solving for the
Vasiliev master fields. More importantly it allows one observing a
certain shift symmetry present in every individual contribution of
the reduced HS vertex. This symmetry being implicit in the
presence of $z$-dominated terms restores manifestly at the level
of the full $z$-independent vertex.

The role of operator $\ord_{-\infty}$ is in rebuilding the
$z$-independent HS vertex out of the reduced one. From the point
of view of the original HS contributions that contain everything
the Vasiliev equations bring, $\ord_{-\infty}$ carries out a
suitable reshuffling that reveals two important structures. One is
a peculiar factorized form of each individual contribution with
respect to a certain product and another one is a realization of
the shift symmetry in terms of objects that enter the factorized
expressions. Both structures are being the building blocks of our
proof of the $z$ -- dominance conjecture. In other words, operator
$\ord_{-\infty}$ can be thought of as something that re-arranges
HS vertex so as to make it free from certain idle terms of various
contributions which sum to zero. In particular, those idle terms
may break the otherwise manifest shift symmetry. This way we get
to our starting point of the analysis that uses expressions
\eqref{Start} as an individual contribution to HS vertex.

%Note, that while we have no proof that eq. \eqref{Start} holds to
%any order $\go C^n$ within the Vasiliev equations, it does hold
%for $n\leq 3$ as clearly follows from \cite{Didenko:2018fgx},
%\cite{Didenko:2020bxd} \cite{Didenko:2019xzz}. So, we keep this
%assumption for all $n$, which however makes sense in view of
%\cite{Didenko:2022qga}, where it indeed was shown to be true to
%all orders within a different approach.

The next step is we examine the $z$ -- independence condition of
our generic expression of a vertex. Analyzing condition
$\dr_z\Upsilon=0$ we arrive at eq. \eqref{zInd1} which is
equivalent to an infinite set of equations for each power of $z$.
Presence of shift symmetry makes them all satisfied provided a
single condition at $z=0$ holds.

One of the central observation of our paper is the so-called {\it
division formula} \eqref{division} which decomposes any function
into two pieces  -- the one which if plugged into a HS vertex
contribution leads to a manifestly $z$-independent expression and
another one called the {\it remainder} which does not. Remarkably,
the form of the remainder is almost identical (up to an
integration measure) to the $z$ -- independence condition. It is
not hard to show that the $z$ -- dominance conjecture fulfills for
a vanishing sum of all remainders. So, a further line of thought
is to prove that it does vanish. More specifically, the idea is to
show that the $z$ -- independence condition perhaps supplemented
with an extra assumption should result in the vanishing of the
remainder contributions.

The extra requirements we assume are the shift symmetry and the
Lorentz invariance. Noting that each of the reduced vertex
contribution considered earlier in
\cite{Didenko:2018fgx},\cite{Didenko:2020bxd},\cite{Didenko:2019xzz}
is shift-symmetric at least up to order $C^3$, we assume it is so
at higher orders as well. Having this symmetry we are able then to
write out a differential equation for any possible vertex that
inherits such a symmetry and analyze its solution space. It turns
out that given the $z$-independence and Lorentz invariance imposed
there are no analytic solutions for the remainder. This fact
brings us to a conclusion that the remainder which is by
construction an analytic function of $y$ should be equal to zero.
This will conclude the proof of the $z$ -- dominance conjecture.

\subsubsection*{Structure of the paper} Since the paper is rather
technical the aim of this small section is to provide reader with some
sort of a guiding map into technical steps of our proof of the $z$
-- dominance conjecture.
\begin{itemize}
\item In Section \ref{Restore} we provide a procedure which allows
one restoring HS vertices from the reduced contributions by virtue
of $\ord_{-\infty}$ operator. The main result of this section is
formula \eqref{result} which prescribes how a generic
$z$-dominated contribution \eqref{generic} transforms to lead to a
consistent vertex.

\item In Section \ref{decZ} we show how one can simplify the
analysis without losing generality by choosing the polynomial in
$z$ pre-exponential of \eqref{contr} to be the linear one. In the
end of this section we provide eq. \eqref{Start} which is the main
subject of the research in this paper.

\item In Section \ref{ZinCond} the $z$-independence restriction
imposed on \eqref{Start} is analyzed leading to $z$-independence
condition \eqref{zInd}. In Section \ref{vertComp} we further find
a class of solutions of \eqref{zInd}.

\item In Section \ref{divS} we prove the division formula
\eqref{division1} which is motivated by the above class. The
meaning of the division formula is in decomposition of a given
function into a part which manifestly solves $z$-independence
condition \eqref{zInd} and the remainder.

\item In Section \ref{PLTS} we provide the notion of the shift
symmetry. This symmetry of the vertex is inherited from each
individual contribution it is composed of. Thus expression
\eqref{Start} obtained in the previous steps should also enjoy
this property.

\item Section \ref{RemZ} shows that the presence of the shift
symmetry leads to a certain differential equation for the
remainder piece.  Its solutions subject to the $z$-independence
condition along with the Lorentz invariance are found and shown to
be non-analytic thus proving that the remainder from division
formula \eqref{division1} equals zero.

\item In Section \ref{Locality} we show how the absence of the
remainder leads to the locality of the vertex. We also propose a
class of local shift-symmetric field redefinitions that arguably
preserve locality of HS vertices.

\item Finally, in Appendices A and B the explicit details for
particular $C^2$ and $C^3$ vertices are presented. These examples
illustrate how the $z$-dominance tools apply in practice.

\end{itemize}

\section{HS vertex and $\ord_\beta$ -- operator}\label{Restore}

As it was already mentioned in the introduction, in solving of the
Vasiliev equations one can not discard any particular vertex
contribution without ruining consistency. However the dominated
terms were omitted in \cite{Didenko:2019xzz} for example.
Justification of this step is the following. Vasiliev system
guarantees that the full vertex is $z$-independent and therefore
one can act on it with any projector on $\dr_z$-cohomologies. Such
action on $z$-independent functions is equivalent to
multiplication by 1. An example of the projector is
\begin{equation}\label{proj}
h_\beta f(z,y)=\ff{1}{(2\pi)^2}\int\dr^2 u\,\dr^2 v\, e^{iu_\alpha
v^\alpha} f(u,y-\beta v)\,.
\end{equation}
Its important feature is it acts trivially on the $z$-dominated
terms in $\beta \rightarrow -\infty $ limit
\cite{Gelfond:2019tac}. Hence an expression with the omitted
dominated terms provides all necessary information of the full HS
vertex and in \cite{Gelfond:2021two} one of the third order in $C$ vertices was
restored from this sort of data. However for our proceeding
analysis the full expression is required since we are going to
study the $z$-independence condition of the vertex. While it is
maybe not clear what to do with the 'reduced vertex' with the
dominated terms dropped off, luckily there is a way to restore the
full one.

Consider an operator introduced in \cite{Didenko:2019xzz}
\begin{equation}
O_\beta f(z,y\vert\, \theta)=\int \frac{\dr^2 u\, \dr^2
v}{(2\pi)^2}e^{iu_\alpha v^\alpha} f(z+u,y-\beta v\vert\,
\theta)\,,
\end{equation}
where $\gb$ is an arbitrary number. So defined projector has the
meaning of the star-product reordering\footnote{See also
\cite{DeFilippi:2021xon} for the analysis of different orderings
in higher-spin context.} and was first defined in
\cite{DeFilippi:2019jqq} for $\gb=1$. By rescaling $z$ and
$\theta$ one modifies this operator\footnote{which is no longer
the reordering operator since it stretches variable $z$.} as
follows (see also \cite{Didenko:2022qga})
\begin{equation}
\ord_\beta f(z,y\vert\, \theta)=\int \frac{\dr^2 u\, \dr^2
v}{(2\pi)^2}e^{iu_\alpha v^\alpha}
f\left(\left(1-\beta\right)z+u,y-\beta v\vert\,
\left(1-\beta\right)\theta\right)\,.
\end{equation}
Remarkable property of operator $\ord_\beta$ is that it acts
trivially on the functions of the specific type in $\beta
\rightarrow -\infty$ limit. Consider a function of the following
form
\begin{equation}\label{class}
f(z,y)=\int_0^1 \dr\mathcal{T}\, e^{i\mathcal{T}z_\alpha
y^\alpha}\phi(\mathcal{T}z,y\vert\, \mathcal{T})\,.
\end{equation}
If there exists such $\epsilon>0$ that
\begin{equation}\label{limit}
\lim_{\mathcal{T}\rightarrow 0}\, \mathcal{T}^{1-\epsilon}
\phi(w,u\vert\, \mathcal{T})=0
\end{equation}
then $\ord_{-\infty}$ act on this function trivially, i.e.
\begin{equation}
\ord_{-\infty} f(z,y)=0\,.
\end{equation}
Expression \eqref{class} that satisfies \eqref{limit} provides the
space of function annihilated by projector $h_\beta$ \eqref{proj}
in $\beta\rightarrow -\infty$ (see
\cite{Gelfond:2019tac},\cite{Didenko:2019xzz}). Hence condition
\eqref{Ex} for example can be relaxed and in practical
computations should be fulfilled only in some neighbourhood of
$\mathcal{T}=0$. Operator $\ord_{-\infty}$ for obvious reasons
leaves $z$-independent functions invariant and enjoys the
following remarkable projection property
\begin{equation}
\ord_{-\infty}\ord_{-\infty}=\ord_{-\infty}
\end{equation}
if applied to HS contributions within the proper functional class
\cite{Gelfond:2019tac} (see Appendix B for $C^3$ example). Hence
one can act with $\ord_{-\infty}$ on expressions with the
dominated terms dropped off (called the reduced ones) and restore
this way the full $z$-independent vertex. Below we are going to
demonstrate how it works in practice.

Every individual contribution obtained in \cite{Didenko:2020bxd}
has the following form
\begin{equation}\label{generic}
f(z,y,t,p_j)=\int_0^1 \dr\mathcal{T} \int \dr^n \rho \, \;
\frac{\pi(\mathcal{T}z,y,t,p_j\vert\,
\rho)}{\mathcal{T}}\exp\Big\{i\mathcal{T}z_\alpha
(y-P_z)^\alpha-iy_\alpha P_y {}^\alpha-it_\alpha
P_t^\alpha\Big\}\omega C^3.
\end{equation}
Here $\int \dr\rho$ stands for all integrations different from
$\mathcal{T}$, while $P_y$, $P_t$, $P_z$ are the $\rho$-dependent
derivatives of fields $\omega$ and $C$, i.e.
\begin{equation}\label{PZ}
P_z(t,p_j|\,\rho)=B_t(\rho) t+B_1(\rho)p_1+\ldots B_3(\rho)p_3\,,
\end{equation}
\begin{equation}\label{PY}
P_y(t,p_j|\,\rho)=A^y_t(\rho)t+A^y_1(\rho)p_1+\ldots
+A^y_3(\rho)p_3\,,
\end{equation}
\begin{equation}\label{PT}
P_t(t,p_j|\,\rho)=A^t_1(\rho)p_1+\ldots +A^t_3(\rho)p_3\,.
\end{equation}
Particular ordering of fields $\omega$ and $C$ is not important as
one can check that contributions to any ordering has the form
\eqref{generic}. For brevity from now on we omit $\omega$ and
$C$'s in the proceeding formulas. Pre-exponential $\pi$ has now no
implicit dependence on $\mathcal{T}$ as compared to the formulas
in the introduction. This is due to the fact that this dependence
is simply $\frac{1}{\mathcal{T}}$ for all contributions and now it
is written explicitly. Note there are no divergences in
\eqref{generic} in integration over $\mathcal{T}$ since
$\pi(\mathcal{T}z,y,\partial\vert\, \rho)$ starts with the first
power in $\mathcal{T} z$ or higher, i.e.
\begin{equation}
\pi(\mathcal{T}z,y,t,p_j \vert\, \rho)=\mathcal{T} z_{\alpha}
\pi^\alpha(\mathcal{T}z,y,t,p_j \vert\, \rho).
\end{equation}
Action of $\ord_\beta$ (before taking the limit) on
\eqref{generic} gives the following
\begin{multline}
\ord_\beta f(z,y,t,p_j)=\ff{1}{(2\pi)^2}\int \dr^2u\, \dr^2 v\, e^{iuv}
\int \dr^n \rho \int_0^{\frac{(1-\beta)\varepsilon}{1-\beta\varepsilon}}
d\mathcal{T}\; \frac{\left(1-\left(\frac{-\beta}{1-\beta}\right)\mathcal{T}\right)}{\mathcal{T}}\times\\
\times \pi\left(\mathcal{T} z-\left(\frac{-\beta}{1-\beta}\right)\mathcal{T} u-
\left(\frac{-\beta}{1-\beta}\right)\mathcal{T} P_y,\left(1-\left(\frac{-\beta}{1-\beta}\right)
\mathcal{T}\right)y-v+\left(\frac{-\beta}{1-\beta}\right)\mathcal{T} P_z,t,p_j\Bigg|\, \rho\right)\times\\
\times\exp\Bigg\{i\mathcal{T} z_\alpha
(y-P_z)^\alpha-i\left(1-\left(\frac{-\beta}{1-\beta}\right)\mathcal{T}\right)
y_\alpha P_y
{}^\alpha-i\left(\frac{-\beta}{1-\beta}\right)\mathcal{T}
P_{z\alpha}P_y {}^\alpha +it^\alpha P_{t\alpha}\Bigg\}\,.
\end{multline}
Taking the limit $\beta \rightarrow -\infty$ one obtains
\begin{multline}
\ord_{-\infty}f(z,y,t,p_j)=\\
=\ff{1}{(2\pi)^2}\int \dr^2u\, \dr^2 v\, e^{iuv}\int d^n \rho \int_0^1 d\mathcal{T}\,
\frac{1-\mathcal{T}}{\mathcal{T}} \pi\Big(\mathcal{T}z-\mathcal{T}u-\mathcal{T}P_y,\;
(1-\mathcal{T})y-v+\mathcal{T}P_z,t,p_j\Big|\, \rho\Big)\times\\
\exp\Big\{i\mathcal{T}z_\alpha(y-P_z)^\alpha-i(1-\mathcal{T})y_\alpha
P_y {}^\alpha-i\mathcal{T}P_{z\alpha} P_y {}^\alpha+it^\alpha
P_{t\alpha}\Big\}\,.
\end{multline}
Shuffling the $u$, $v$ integration variables it can be brought to
a form
\begin{multline}\label{Oinf}
\ord_{-\infty}f(z,y,t,p_j)=\\
=\ff{1}{(2\pi)^2}\int \dr^2u\, \dr^2 v\, e^{iuv}\int \dr^n \rho \int_0^1 d\mathcal{T}\,
\frac{1-\mathcal{T}}{\mathcal{T}} \pi\Big(\mathcal{T}z-\mathcal{T}v-\mathcal{T}P_y,\;
(1-\mathcal{T})y+u+\mathcal{T}P_z,\; t,p_j\Big|\, \rho\Big)\times\\
\exp\Big\{i\mathcal{T}z_\alpha(y-P_z)^\alpha-i(1-\mathcal{T})y_\alpha
P_y {}^\alpha-i\mathcal{T}P_{z\alpha} P_y {}^\alpha+it^\alpha
P_{t\alpha}\Big\}\,.
\end{multline}
Function $\pi$ is polynomial in $z$ and can be written down as a
sum of various monomials
\begin{equation}\label{sep}
\pi(\mathcal{T}z,y,t,p_j \vert\, \rho)=\mathcal{T}^n
z_{\alpha_1}\ldots z_{\alpha_n} \pi^{\alpha_1\ldots
\alpha_n}(y,t,p_j \vert\, \rho)
\end{equation}
Consider any term from this sum and introduce the following
$\circledast$ operation (see also \cite{Didenko:2022qga})
\begin{equation}\label{halfP}
f(y)\circledast \Gamma(z,y):=\ff{1}{(2\pi)^2}\int \dr^2u\, \dr^2
v\, e^{iuv}\, f(y+u)\Gamma(z-v,y)\,,
\end{equation}
then the previous expression can be rewritten as
\begin{equation}\label{result}
\ord_{-\infty}f(z,y,t,p_j) =\int \dr^n \rho\,   e^{+iy^\alpha P_y
{}_\alpha+it^\alpha P_{t\alpha}}\pi^{\alpha(n)}(y,\partial\vert
\rho)\circledast\int_0^1
\dr\mathcal{T}\frac{1-\mathcal{T}}{\mathcal{T}} (\mathcal{T}
z_{\alpha})^n e^{i\mathcal{T}z_\alpha(y-P_z)^\alpha}\,.
\end{equation}

\section{Reducing pre-exponential}\label{decZ}
The higher the order of perturbation theory, the higher the power
of $z$ that appears in the pre-exponential of \eqref{result}. It
growth is estimated as $z^{N-1}$, where $N$ is the order of
perturbations in $C$. Just as in \cite{Gelfond:2021two} for
simplification of the proceeding analysis one can systematically
decrease the power of $z$ in the pre-exponential up to manifestly
local $\dr_z$-cohomologies. The best one can do without facing any
divergencies is a linear in $z$ pre-exponential . This procedure
is easier  to perform in practice with the $z$-dominated terms
dropped off. Suppose one has the expression
\begin{equation}\label{toUniform}
\int_0^1 \dr\mathcal{T} \int \dr^n \rho \, R(\rho) \,
\frac{\mathcal{T}z_\alpha \xi^\alpha \mathcal{T} z_\beta
\zeta^\beta}{\mathcal{T}}\exp\big\{i\mathcal{T}z_\alpha
(y-P_z)^\alpha-iy_\alpha P_y {}^\alpha-it_\alpha
P_t^\alpha\big\}\,.
\end{equation}
Here $R(\rho)$ is some function of $\rho$, while $\xi$ and $\zeta$
are any spinors from the set $\{y,t,p_j\}$. We want to treat every
contribution in a uniform way and for this purpose we introduce
the additional integration variables along with the corresponding
delta-functions of the form
\begin{equation}\label{dPz}
\int \dr\varrho^z_t\, \delta(\varrho^z_t-B_t(\rho))\int
\dr\varrho_1^z\, \delta(\varrho^z_1-B_1(\rho))\ldots \int
\dr\varrho_3^z\, \delta(\varrho^z_3-B_3(\rho))\,,
\end{equation}
\begin{equation}\label{dPy}
\int \dr\varrho^y_t\, \delta(\varrho^y_t-A^y_t(\rho))\int
\dr\varrho_1^y\, \delta(\varrho^y_1-A_1^y(\rho))\ldots \int
\dr\varrho_3^y\, \delta(\delta(\varrho^y_3-A_3^y(\rho))\,,
\end{equation}
\begin{equation}\label{dPt}
\int \dr\varrho_1^t\,\delta(\varrho^t_1-A^t_1(\rho))\ldots \int
\dr\varrho_3^t\,\delta(\varrho^t_3-A_3^t(\rho))\,.
\end{equation}
Then expression \eqref{toUniform} casts into
\begin{equation}
\int_0^1 \dr\mathcal{T} \int \dr^n \rho \int \dr\varrho\,
\mathcal{R}(\rho,\varrho) \, \frac{\mathcal{T}z_\alpha \xi^\alpha
\mathcal{T} z_\beta
\zeta^\beta}{\mathcal{T}}\exp\Big\{i\mathcal{T}z_\alpha
(y-P_z)^\alpha-iy_\alpha P_y {}^\alpha-it_\alpha
P_t^\alpha\Big\}\,.
\end{equation}
Here $\int \dr \varrho$ is the integral over all $\varrho^\mu_\nu$
while the new measure $\mathcal{R}$ absorbs all the delta-function
from \eqref{dPz},\eqref{dPy},\eqref{dPt}. We partially perform the
integration so that only $\varrho$ are present in
\eqref{PZ},\eqref{PY}, \eqref{PT}, i.e
\begin{equation}
P_z(t,p_j|\,\rho)=\varrho^z_t t+\varrho^z_1 p_1+\ldots \varrho^z_3
p_3\,,
\end{equation}
same is for $P_y$ and $P_t$. Now let $\xi$ be either $t$ or $p_j$
then any such term in the pre-exponential can be represented as
the derivative of the exponential over corresponding $\varrho$.
For example let $\xi^\alpha=t^\alpha$ then
\begin{multline}\label{zt}
\int_0^1 \dr\mathcal{T} \int \dr^n \rho \int \dr\varrho\, \mathcal{R}(\rho,\varrho) \, \frac{\mathcal{T}z_\alpha t^\alpha \mathcal{T} z_\beta \zeta^\beta}{\mathcal{T}}\exp\Big\{i\mathcal{T}z_\alpha (y-P_z)^\alpha-iy_\alpha P_y {}^\alpha-it_\alpha P_t^\alpha\Big\}=\\
=\int_0^1 \dr\mathcal{T} \int \dr^n \rho \int \dr\varrho\, \mathcal{R}(\rho,\varrho) \, \frac{\mathcal{T} z_\beta \zeta^\beta}{\mathcal{T}}i\frac{\partial}{\partial \varrho^z_t}\exp\Big\{i\mathcal{T}z_\alpha (y-P_z)^\alpha-iy_\alpha P_y {}^\alpha-it_\alpha P_t^\alpha\Big\}=\\
=-i\int_0^1 \dr\mathcal{T} \int \dr^n \rho \int \dr\varrho\,
\frac{\partial}{\partial
\varrho^z_t}\Big(\mathcal{R}(\rho,\varrho)\Big) \,
\frac{\mathcal{T} z_\beta
\zeta^\beta}{\mathcal{T}}\exp\Big\{i\mathcal{T}z_\alpha
(y-P_z)^\alpha-iy_\alpha P_y {}^\alpha-it_\alpha
P_t^\alpha\Big\}\,.
\end{multline}
If $\xi$ equals $y$ then we can add and subtract $P_z$ and then
integrate by parts over $\mathcal{T}$, i.e.
\begin{multline}\label{zy}
\int_0^1 \dr\mathcal{T} \int \dr^n \rho \int \dr\varrho\, \mathcal{R}(\rho,\varrho) \, \frac{\mathcal{T}z_\alpha (y-P_z)^\alpha \mathcal{T} z_\beta \zeta^\beta}{\mathcal{T}}\exp\Big\{i\mathcal{T}z_\alpha (y-P_z)^\alpha-iy_\alpha P_y {}^\alpha-it_\alpha P_t^\alpha\Big\}+\\
+\int_0^1 \dr\mathcal{T} \int \dr^n \rho \int \dr\varrho\,
\mathcal{R}(\rho,\varrho) \, \frac{\mathcal{T}z_\alpha
P_z{}^\alpha \mathcal{T} z_\beta
\zeta^\beta}{\mathcal{T}}\exp\Big\{i\mathcal{T}z_\alpha
(y-P_z)^\alpha-iy_\alpha P_y {}^\alpha-it_\alpha
P_t^\alpha\Big\}\,.
\end{multline}
Here the first term can be realized as the derivative over
$\mathcal{T}$ and after partial integration it brings local
$\dr_z$-cohomology term\footnote{There are two boundary terms: one
is indeed $\dr_z$-cohomology when $\mathcal{T}=0$ while the other
one can be discarded since it can be treated as a function that
fulfils \eqref{limit}}. The second term can be realized in terms
of derivatives over various $\rho^z_\nu$ like in \eqref{zt}.
Hence, all the contributions can be brought to the linear in $z$
pre-exponential. Note, the top power of $z$ being two is not an
obstruction as one can decrease the $z\dots z$-monomial to the
linear one for any power $n$.

Now we have contributions that are linear in $z$ in the
pre-exponential modulo local cohomology terms which we are not
going to consider since those are manifestly local already. The
sum of all the leftover terms looks as follows
\begin{equation}\label{SUM}
 \int_0^1 \dr\mathcal{T}\int \dr^n\rho\int \dr\varrho\, \sum_k\mathcal{R}_k(\rho,\varrho)
 z_\alpha \pi^\alpha_k(y,t,p_j)\exp\Big\{i\mathcal{T}z_\alpha (y-P_z)^\alpha-iy_\alpha P_y {}^\alpha-it_\alpha
 P_t^\alpha\Big\}\,.
\end{equation}
We are not going to distinguish between $\rho$ and $\varrho$
anymore\footnote{One can simply rename various $\varrho^\mu_\nu$ as
$\rho_{n+1},\; \rho_{n+2},\ldots$.} for brevity, as we write the
integral over all these variables simply as
\begin{equation}
\int \mathscr{D}\rho:=\int \dr^n\rho \int \dr\varrho\,,
\end{equation}
also in what follows we are going to use the following shorthand
notation
\begin{equation}\label{PI}
\Pi^\alpha(y,t,p_j|\,\rho):=\sum_k\mathcal{R}_k(\rho,\varrho)
\pi^\alpha_k(y,t,p_j)\exp\Big\{-iy_\alpha P_y {}^\alpha-it_\alpha
P_t^\alpha\Big\}\,,
\end{equation}
which is manifestly local provided $\pi$ contains no infinite
$pp$-contractions . With the help of the new notation
\eqref{halfP} the expression for the vertex modulo manifestly
local $\dr_z$-cohomologies looks as follows
\begin{equation}\label{Start}
\Upsilon(y,t,p_j)=\int \mathscr{D}\rho\,
\Pi^\al(y,t,p_j|\,\rho)\circledast\int_0^1\dr\mathcal{T}\,
(1-\mathcal{T})z_\al\, e^{i\mathcal{T}z_\alpha(y-P_z)^\alpha}\,.
\end{equation}
Here we applied $\ord_{-\infty}$ operator to \eqref{SUM}. Formula
\eqref{Start} is the main subject for the proceeding analysis. At
this stage of function $\Pi^{\al}$ we only assume that it contains
no nonlocal $pp$-contractions which is equivalent to the statement
that non-localities of the vertex contribution are $z$-dominated.
This of course says nothing at all about the locality behavior of
$\Upsilon$ itself. In practice \eqref{Start} has always apparent
nonlocalities due to $\circledast$ -- operation that gives rise to
the exponentiated $pp$ -- contractions.

\section{$Z$-independence condition}\label{ZinCond}
As it was already mentioned in the introduction the Vasiliev
system guarantees that HS vertices are $z$-independent, hence
\eqref{Start} should be $z$-independent too. Straightforward
computation yields
\begin{multline}
\frac{\partial}{\partial z^\alpha}\Bigg(\int \mathscr{D}\rho\, \Pi^\beta(y,t,p_j|\,\rho)\circledast\int_0^1\dr \mathcal{T}\, \frac{1-\mathcal{T}}{\mathcal{T}}\mathcal{T}z_\beta\, e^{i\mathcal{T}z_\alpha(y-P_z)^\alpha}\Bigg)=\\
=\int \mathscr{D} \rho \, \Pi_\alpha(y,t,p_j\vert\, \rho)\circledast\int_0^1 \dr\mathcal{T}\, \mathcal{T}\, e^{i\mathcal{T}z_\alpha (y-P_z)^\alpha}+\\
+\int \mathscr{D} \rho \, \Pi^\beta(y,t,p_j\vert\,
\rho)\circledast\int_0^1 \dr\mathcal{T}\,
(1-\mathcal{T})i\mathcal{T}z_\alpha(y-P_z)_\beta
e^{i\mathcal{T}z_\alpha (y-P_z)^\alpha}\,.
\end{multline}
Then plugging \eqref{halfP} explicitly and representing the first
term as
\begin{equation}
\Pi_\alpha=\Pi^\beta \epsilon_{\beta \alpha}=-\Pi^\beta
\frac{\partial(z-v)_\alpha}{\partial v^\beta}
\end{equation}
and then integrating by parts with respect to $v$ one can bring
the $z$-independence condition to the following form
\begin{equation}\label{zInd1}
\int \mathscr{D} \rho \, \Pi^\beta(y,t,p_j\vert\,
\rho)(y-P_z)_\beta \circledast \int_0^1 \dr\mathcal{T} \,
\mathcal{T}z_\alpha e^{i\mathcal{T}z_\alpha (y-P_z)^\alpha}=0\,.
\end{equation}
Eq. \eqref{zInd1} should be fulfilled for any value of $z$ upon
$\circledast$ - computation . Therefore, it amounts to an infinite
chain of conditions assigned to any given power of $z$ in its
Taylor series. Particularly it should be true for $z=0$ leading to
\begin{equation}\label{zInd}
\frac{\partial}{\partial y^\alpha}\int_0^1 \dr\mathcal{T} \int
\mathscr{D}\rho\, \mathcal{T}(y-P_z)^\beta \Pi_\beta
\big((1-\mathcal{T})(y-P_z)+P_z,t,p_j|\,\rho\big)=0\,.
\end{equation}
Eq. \eqref{zInd1} can be further massaged using \eqref{halfP} to
\be\label{zInd2}
\int_0^1 \dr\mathcal{T} \int \mathscr{D}\rho\,
\mathcal{T}\p_{\al}\left(e^{-iz_{\gb}x^{\gb}}f(x)\right)=0\,,
\ee
where
\be\label{f}
f(y; t, p_i|\,\rho)=(y-P_z)^{\al}\Pi_\al(y,t,p_j|\,\rho)\,,\quad
x=(1-\mathcal{T})y+\mathcal{T}P_z\,,\quad \p_{\al}=\ff{\p}{\p
x^{\al}}\,.
\ee
Since $z$ is arbitrary in \eqref{zInd2} and \eqref{zInd2} must
hold for each power of $z$ one can replace exponential $e^{-iz_\alpha x^\alpha}$
with arbitrary function $\psi(x)$
\be\label{zInd3}
\int_0^1 \dr\mathcal{T} \int \mathscr{D}\rho\,
\mathcal{T}\p_{\al}\left(\psi(x)f(x; t,p_i|\,\rho)\right)=0\,.
\ee
Therefore, $f(y)$ as defined in \eqref{f} should be such that
\eqref{zInd3} is fulfilled for any function $\psi(y)$.

\section{Manifest $z$-independence}\label{vertComp}
Eq. \eqref{Start} leads to a manifestly $z$-independent expression
if function $\Pi^\beta(y,t,p_j|\,\rho)$ is of the following form
\begin{equation}\label{div}
\Pi^\beta(y,t,p_j|\,\rho)=(y-P_z)^\beta \Pi(y,t,p_j|\,\rho)\,.
\end{equation}
Indeed, rewriting \eqref{Start} explicitly one has
\begin{multline}\label{preVer}
\ff{1}{(2\pi)^2}\int \mathscr{D}\rho \int \dr^2u\, \dr^2 v\,
e^{iuv}\, \Pi(y+u,t,p_j|\,\rho)\int_0^1 \dr\mathcal{T}\,
(1-\mathcal{T})\Big((z-v)_\beta
(y-P_z)^\beta-\\-u_\beta(z-v)^\beta\Big)e^{i\mathcal{T}(z-v)_\alpha
(y-P_z)^\alpha}\,.
\end{multline}
Integrating then by parts over $u$ and $v$ and using
representation as the derivative with respect to $\mathcal{T}$
\begin{equation}
(z-v)_\beta (y-P_z)^\beta e^{i\mathcal{T}(z-v)_\alpha
(y-P_z)^\alpha}=-i\frac{\partial}{\partial \mathcal{T}}
e^{i\mathcal{T} (z-v)_\alpha (y-P_z)^\alpha}
\end{equation}
expression \eqref{preVer} turns into
\begin{equation}
\ff{1}{(2\pi)^2}\int\mathscr{D}\rho \int \dr^2u\, \dr^2 v\,
e^{iuv}\, \Pi(y+u,t,p_j|\,\rho)\int_0^1 \dr\mathcal{T}\,
(1-\mathcal{T})\Big(\underbrace{-i\frac{\partial}{\partial
\mathcal{T}}}+2i+i\mathcal{T}\frac{\partial}{\partial
\mathcal{T}}\Big)e^{i\mathcal{T}(z-v)_\alpha (y-P_z)^\alpha}\,.
\end{equation}
Here the underbraced term results in $\dr_z$-cohomology, while the
other terms cancel each other and hence the vertex in terms of
function $\Pi$ is simply
\begin{equation}\label{Vertex}
\Upsilon(y,t,p_j)=i\int\mathscr{D}\rho\, \Pi(y,t,p_j|\,\rho)\,.
\end{equation}
We will refer to $\Pi$ as to the {\it vertex function} in what
follows. This suggests the mechanism how HS vertex gets manifestly
$z$-independent. The idea is the $z$-independence condition
\eqref{zInd1} may resolve function $\Pi^{\al}$ in the form
\eqref{div} up to the $\rho$-integration. To check if it is really
so or not we are going to bring $\Pi^\beta(y,t,p_j)$ to the form
\eqref{div} up to the remainder, i.e. {\it divide}
$\Pi^\beta(y,t,p_j)$ over $(y-P_z)^\beta$.

\section{Division formula}\label{divS}
In this section we elaborate on the procedure that allows one to
write $\Pi^\beta(y,t,p_j|\,\rho)$ as \eqref{div} up to a
remainder. Consider the following identity
\begin{equation}\label{Piform}
\Pi^{\beta}(y,\partial\vert\, \rho)=\int \dr\xi\,
\theta(\xi)\delta(1-\xi)r(\xi)\Pi^\beta\big(\xi(y-P_z)+P_z,t,p_j\vert\,
\rho\big)\,,
\end{equation}
provided $r(1)=1$. Rewriting $\delta(1-\xi)$ as
$\frac{\partial}{\partial \xi}\theta(1-\xi)$ and integrating by
parts one obtains
\begin{multline}\label{div2}
\Pi_{\beta}(y,t,p_j\vert\, \rho)=\int \dr\xi\, \delta(\xi)r(\xi)\theta(1-\xi) \,\Pi_{\beta}\big(\xi(y-P_z)+P_z,t,p_j\vert\, \rho\big)+\\
\int_0^1 \dr\xi\,\frac{\partial r(\xi)}{\partial \xi}
\Pi_{\beta}\big(\xi(y-P_z)+P_z,t,p_j\vert\,
\rho\big)+\int_0^1\dr\xi\, r(\xi) \frac{\partial}{\partial \xi}
\Pi_{\beta}\big(\xi(y-P_z)+P_z,t,p_j\vert\, \rho\big)\,.
\end{multline}
The derivative over $\xi$ in the last term can be taken and with
the help of the Schouten identities brought to a form
\begin{multline}
\frac{\partial}{\partial \xi}  \Pi_{\beta}\big(\xi(y-P_z)+P_z,t,p_j\vert\, \rho\big)=(y-P_z)^\sigma
\frac{\partial}{\partial \big(\xi(y-P_z)+P_z\big)^\sigma}  \Pi_{\beta}\big(\xi(y-P_z)+P_z,t,p_j\vert\, \rho\big)=\\
=(y-P_z)_\beta \frac{1}{\xi}\frac{\partial}{\partial y^\sigma}  \Pi^{\sigma}\big(\xi(y-P_z)+P_z,t,p_j\vert\, \rho\big)
+\frac{1}{\xi}\frac{\partial}{\partial y^\beta}(y-P_z)^\sigma \Pi_\sigma \big(\xi(y-P_z)+P_z,t,p_j\vert\, \rho\big)-\\
-\frac{1}{\xi}\Pi_{\alpha}\beta\big(\xi(y-P_z)+P_z,t,p_j|\,\rho\big)\,.
\end{multline}
Plugging the expression for the derivative back into \eqref{div2}
we obtain
\begin{multline}
\Pi_{\beta}(y,t,p_j\vert\, \rho)=r(0)\Pi_\beta (P_z,t,p_j|\,\rho)+
(y-P_z)_\beta \int_0^1 \dr\xi \, \frac{r(\xi)}{\xi} \frac{\partial}{\partial y^\sigma}\Pi^\sigma\big(\xi(y-P_z)+P_z,t,p_j|\,\rho\big)+\\
+\frac{\partial}{\partial y^\beta}\int_0^1 \dr\xi \, \frac{r(\xi)}{\xi}(y-P_z)^\sigma\Pi_{\sigma}\big(\xi(y-P_z)+P_z,t,p_j\vert\, \rho\big)+\\
+\int_0^1
\dr\xi\Big(r^\prime(\xi)-\frac{r(\xi)}{\xi}\Big)\Pi_{\beta}\big(\xi(y-P_z)+P_z,t,p_j\vert\,
\rho\big)\,.
\end{multline}
We want the last term to vanish thus we take
\begin{equation}
r(\xi)=\xi\,.
\end{equation}
With this specific choice for $r(\xi)$ the identity we started
from acquires the form
\begin{multline}\label{division}
\Pi_\sigma(y,t,p_j|\,\rho)\equiv (y-P_z)_\sigma \int_0^1 \dr\xi\, \frac{\partial}{\partial (y-P_z)^\beta} \Pi^\beta\big(\xi(y-P_z)+P_z,t,p_j|\,\rho\big)+\\
+\frac{\partial}{\partial y^\sigma}\int_0^1 \dr\xi \,
(y-P_z)^\beta \Pi_\beta\big(\xi(y-P_z)+P_z,t,p_j|\,\rho\big)\,.
\end{multline}
Here the first term is a realization of $\Pi_{\sigma}$ as in
\eqref{div} in terms of the vertex function
\be\label{verfunc}
\Pi(y,t,p_j|\,\rho)=\int_0^1 \dr\xi\, \frac{\partial}{\partial
(y-P_z)^\beta} \Pi^\beta\big(\xi(y-P_z)+P_z,t,p_j|\,\rho\big)\,,
\ee
but there is an additional term which will be referred to as the
{\it remainder} in what follows. Presence of the remainder is not
surprising since \eqref{division} is just an identity and not
every $\Pi_\sigma$ should lead to a $z$-independent expression
when feeded into \eqref{Start} (see section \ref{vertComp}).
However one can plug
\begin{equation}\label{div4}
(y-P_z)_\sigma \int_0^1 \dr\xi\, \frac{\partial}{\partial
(y-P_z)^\beta} \Pi^\beta\big(\xi(y-P_z)+P_z,t,p_j|\,\rho\big)
\end{equation}
into \eqref{Start} in place of $\Pi^\beta(y,t,p_j|\,\rho)$ and
obtain precisely the same result as if the original
$\Pi^\beta(y,t,p_j|\,\rho)$ was in the expression. Indeed since
\eqref{Start} is $z$-independent one can choose any
$\dr_z$-cohomology projector. For our purpose we chose the one that
sets $z$ to zero. Put it differently, the difference that the
remainder part brings is exactly the $z$-independence condition
\eqref{zInd1}. Straightforward computation yields
\begin{equation}\label{Vd}
\Upsilon(y,t,p_j)=i\int\mathscr{D}\rho \int_0^1 \dr\mathcal{T}\,
\frac{\partial}{\partial
y^\sigma}\Pi^\sigma\big((1-\mathcal{T})(y-P_z)+P_z,t,p_j|\,\rho\big)\,,
\end{equation}
which is precisely the one obtained from  \eqref{div4} after
changing the variables $\xi\rightarrow(1-\mathcal{T})$. One might
want to change the variables in the division formula in that
fashion which reads
\begin{multline}\label{division1}
\Pi_\sigma(y,t,p_j|\,\rho)\equiv (y-P_z)_\sigma \int_0^1 \dr\mathcal{T}\,
\frac{\partial}{\partial y^\beta} \Pi^\beta\big((1-\mathcal{T})(y-P_z)+P_z,t,p_j|\,\rho\big)+\\
+\frac{\partial}{\partial y^\sigma}\int_0^1 \dr\mathcal{T} \,
(y-P_z)^\beta
\Pi_\beta\big((1-\mathcal{T})(y-P_z)+P_z,t,p_j|\,\rho\big)\,.
\end{multline}
The last term looks very similar to the integrand from the
$z$-independence condition \eqref{zInd}, but it is not the same!
The difference being the $\mathcal{T}$-measure. To some surprise
the $z$-independence condition is not equivalent to the absence of
the remainder. A precise relation between the remainder and the
$z$-independence condition is given in the proceeding sections.
However, before we continue an illustrative example of the
$z$-cancellation mechanism at the second order zero-form vertex is
provided below.

\subsection{$\Upsilon(\go,C,C)$ vertex}
Consider for example vertex $\Upsilon_{\go C C}$. Its
contributions obtained from the generating system are given in the
Appendix A. They are already linear in $z$ so the expression for
the vertex can be written as \eqref{Start}. Note that $P_z$ is
$\rho$-independent for all second order vertices and is simply
\begin{equation}
P_z=-p_1-p_2-t\, .
\end{equation}
So, we have
\begin{equation}
\Upsilon_{\go CC}=\left(\int \mathscr{D}\rho\,
\Pi^\beta(y,t,p_j|\,\rho)\right)\circledast\int_0^1
\dr\mathcal{T}\, z_\beta e^{i\mathcal{T} z_\alpha
(y-P_z)^\alpha}\,.
\end{equation}
$Z$-inpendence condition \eqref{zInd} leads to\footnote{One can
expect a $y$-independent constant on the right of \eqref{2nd}. It
should be however zero as immediately follows from setting $y=P_z$
on the left.}
\begin{equation}\label{2nd}
(y-P_z)_\beta\int\mathscr{D}\rho\, \Pi^\beta(y,t,p_j|\,\rho)=0\,.
\end{equation}
The latter implies that after integration \eqref{division1} over
$\rho$ the term that corresponds to the remainder vanishes and one
is left with
\begin{equation}\label{2ndO}
\int \mathscr{D}\rho\, \Pi_\sigma(y,t,p_j|\,\rho)= (y-P_z)_\sigma
\int\mathscr{D}\rho \int_0^1 \dr\mathcal{T}\,
\frac{\partial}{\partial y^\beta}
\Pi^\beta\big((1-\mathcal{T})(y-P_z)+P_z,t,p_j|\,\rho\big)\,.
\end{equation}
Note that r.h.s. is the vertex times $(y-P_z)_\sigma$ (see
\eqref{Vd}). Since the l.h.s. is local, equation \eqref{2ndO}
implies that the second order vertex $\Upsilon_{\go CC}$ is also
local. Let us also note that the integral of the vertex function
\be
\Phi(y, t, p_j)=\int\mathscr{D}\rho\, \Pi(y,t,p_j|\,\rho)
\ee
and correspondingly the vertex itself \eqref{Vertex} can be easily
found from \eqref{2nd} in a purely algebraic way. Indeed,
\be\label{2ndver}
\int\mathscr{D}\rho\,
\Pi_\al(y,t,p_j|\,\rho)=(y-P_z)_{\al}\Phi\quad\Rightarrow\quad
\Phi=\ff{t^{\al}\int\mathscr{D}\rho\,
\Pi_\al(y,t,p_j|\,\rho)}{t\cdot(y-P_z)}\,.
\ee
There can be any spinor different from $y-P_z$ in place of the
exterior $t$ in \eqref{2ndver}, which choice gives rise to a
particularly short way to the final $CC$ -- vertex in practice. An
apparent non-analyticity in the form of the final result is
artificial and cancels out in the explicit expressions. Since in
the example above $P_z$ is $\rho$-independent, function $\Phi$ is
found algebraically. Using \eqref{Vertex} one immediately finds
\be
\Upsilon_{C^2}=i\Phi\,.
\ee
Its manifest form and details of derivation we leave for Appendix
A.

The main difficulty in the analysis of the third order vertices
and beyond is due to the fact that $P_z$ no longer
$\rho$-independent and therefore factor $y-P_z$ can not be
separated from the $\rho$-integral.

\section{Shift symmetry}\label{PLTS}
To bring HS vertex to manifestly $z$-independent form may take in
general infinite steps of applying the Schouten identities and
partial integrations. However in practice this process terminates
in few steps. One may expect some symmetry which is responsible
for such a termination. The symmetry that we observe
experimentally at lower orders up to $C^3$ using the results of
\cite{Didenko:2020bxd} amounts to the following transformations
\begin{equation}\label{3O}
\Upsilon_{C^3}(y,t,p_1+a,p_2-a,p_3+a)=e^{ia_\alpha
(t+y)^\alpha}\Upsilon_{C^3}(y,t,p_1,p_2,p_3)\,.
\end{equation}
One also checks that the second order vertices enjoy this symmetry
too, i.e.
\begin{equation}\label{2O}
\Upsilon_{C^2}(y,t,p_1+a,p_2-a)=e^{ia_\alpha
(t+y)^\alpha}\Upsilon_{C^2}(y,t,p_1,p_2)\,.
\end{equation}
While property \eqref{2O} can be observed by simply looking at the
particular given vertex (see \cite{Didenko:2018fgx}, \cite{Didenko:2019xzz}), at order $C^3$
property \eqref{3O} is less clear.

To prove \eqref{3O} one can check that all of the corresponding
vertex contributions from \cite{Didenko:2020bxd} satisfy the
following properties (see \eqref{generic}  for definitions)
\begin{equation}\label{Pza}
P_z(t,p_1+a,p_2-a,p_3+a)=P_z(t,p_1,p_2,p_3)\,,
\end{equation}
\begin{equation}\label{Pya}
P_y(t,p_1+a,p_2-a,p_3+a)=a+P_y(t,p_1,p_2,p_3)\,,
\end{equation}
\begin{equation}\label{Pta}
P_t(p_1+a,p_2-a,p_3+a)=a+P_t(p_1,p_2,p_3)\,,
\end{equation}
\begin{equation}\label{pia}
\pi(y,t,p_1+a,p_2-a,p_3+a)=\pi(y,t,p_1,p_2,p_3)\,.
\end{equation}
Strictly speaking the equalities above should be understood on
integral $\int \dr^n \rho \, {\pi(\mathcal{T}z,y,t,p_j\vert\,
\rho)}$ for \eqref{Pza}-\eqref{Pta} and $\int \dr^n \rho$ for \eqref{pia} although in practice they simply
saturate some $\gd$-functions stored in ${\pi(\mathcal{T}z,y,t,p_j\vert\,
\rho)}$ (see \eqref{START}-\eqref{FINISH}). Thus, modulo dominated terms each
reduced individual contribution enjoys
\begin{equation}
f(z,y,t,p_1+a,p_2-a,p_3+a)=e^{ia_\alpha
(t+y)^\alpha}f(z,y,t,p_1,p_2,p_3)\,.
\end{equation}
This observations correspondingly yields
\begin{equation}\label{PLT3}
\Pi_\sigma(y,p_1+a,p_2-a,p_3+a)=e^{ia_\alpha(t+y)^\alpha}\Pi_\sigma(y,p_1,p_2,p_3)\,,
\end{equation}
on $\int \mathscr{D}\rho$ integral which follows from \eqref{Pza}-\eqref{pia} (recall the definition
of $\Pi_\sigma$ is given by \eqref{PI}). Now, consider a shifted
vertex
\begin{equation}
\Upsilon_{C^3}(y,t,p_1+a,p_2-a,p_3+a)= \int \mathscr{D}\rho\,
e^{ia_\alpha(t+y)^\alpha}
\Pi^\beta(y,t,p_j|\,\rho)\circledast\int_0^1\dr\mathcal{T}\,
\frac{1-\mathcal{T}}{\mathcal{T}}\mathcal{T}z_\beta\,
e^{i\mathcal{T}z_\alpha(y-P_z)^\alpha}\,.
\end{equation}
Writing $\circledast$-product explicitly\eqref{halfP} this expression casts into
\begin{multline}
\ff{1}{(2\pi)^2}\int \dr^2u \dr^2v \, e^{iu_\alpha v^\alpha} \int
\mathscr{D}\rho\, e^{ia_\alpha(t+y+u)^\alpha}
\Pi^\beta(y+u,t,p_j|\,\rho)\int_0^1\dr\mathcal{T}\,
\frac{1-\mathcal{T}}{\mathcal{T}}\mathcal{T}(z_\beta-v_\beta)\,
e^{i\mathcal{T}(z_\alpha-v_\alpha)(y-P_z)^\alpha}=\\
=e^{ia_\alpha (t+y)^\alpha}\int \mathscr{D}\rho\,
 \Pi^\beta(y,t,p_j|\,\rho)\circledast\int_0^1\dr\mathcal{T}\,
\frac{1-\mathcal{T}}{\mathcal{T}}\mathcal{T}(z_\beta-a_\beta)\,
e^{i\mathcal{T}(z_\alpha-a_\alpha)(y-P_z)^\alpha}=\\
=e^{ia_\alpha (t+y)^\alpha}\Upsilon(y, t, p_1, p_2, p_3)\,.
\end{multline}
Here in the last equality we made use of the fact that since
vertex is $z$-independent one can redefine $z-a\to z$ and hence
\eqref{3O} indeed takes place. Note that the decrease in powers of
$z$ in the pre-exponential described in section \ref{decZ} in no
way can spoil the symmetry property. Thus, \eqref{Start} also
transforms accordingly. The shift symmetry observed at first few
orders here in fact holds at any order as has been recently shown
in \cite{Didenko:2022qga}\footnote{The definition of the shift symmetry used in
\cite{Didenko:2022qga} lacks sign alternation as compared to \eqref{3O} due to a
different definition of the physical fields. This difference is
purely a matter of convention.} using a different approach.
Despite the fact that so far one can directly check \eqref{PLT3}
only at orders $C^3$ and $C^2$ within the Vasiliev equations, in
what follows we assume
\be
\Pi_\sigma(y, t, p_1+a, p_2-a,
p_3+a,\ldots)=e^{ia_\alpha(t+y)^\alpha}\Pi_\sigma(y, t, p_1, p_2,
p_3,\ldots)\,,\label{PLT}
\ee
\be
P_z(t, p_1+a, p_2-a, p_3+a,\ldots)=P_z(t, p_1, p_2,
p_3,\ldots)\label{PLT1}
\ee
at any order. If true it allows us to make all order statements on
HS locality. Any HS vertex originates from various contributions
from the generating equations. Each contribution has its own
$\Pi_{\al}$ and $P_{z\, \al}$ which we call {\it shift-symmetric}
if they satisfy \eqref{PLT} and \eqref{PLT1}. The known $C^2$ and
$C^3$ vertex contributions generated by the Vasiliev equations as
well as the all order ones from \cite{Didenko:2022qga} do satisfy
\eqref{PLT}-\eqref{PLT1}.

It is interesting to note how shift symmetry resolves an infinite
chain of the $z$-independence conditions packed in \eqref{zInd1}
by reducing all of them to a single one \eqref{zInd}. Indeed,
consider the $z$-independence condition taken at $z=0$,
\eqref{zInd}. Using now \eqref{PLT} and \eqref{Pza} one arrives at
\be\label{indplt}
e^{ia_{\al}t^{\al}}\frac{\partial}{\partial y^\alpha}\int_0^1
\dr\mathcal{T} \int \mathscr{D}\rho\, \mathcal{T}(y-P_z)^\beta
\Pi_\beta \big(x,t,p_j|\,\rho\big)e^{ia_{\al}x^{\al}}=0\,,
\ee
where $x$ is defined in \eqref{f}. As prefactor $e^{iat}$ cancels
out, eq. \eqref{indplt} gets precisely equivalent to \eqref{zInd2}
upon replacing $a\to -z$. Therefore, the shift symmetry provides a
mechanism that underlies $z$-independence \eqref{zInd2} in
practice.

\section{Remainder equals zero}\label{RemZ}
In the previous section the so-called division formula  which
divides a function into two pieces called the vertex function and
the remainder was introduced. In this section we show that the
remainder vanishes upon integration over $\rho$, provided
$z$-independence \eqref{zInd} and the shift symmetry are imposed.
One should also assume Lorentz invariance by default. To proceed
we introduce two auxiliary functions
\begin{equation}\label{Z}
\mathscr{Z}(p_i):=\int \mathscr{D}\rho \int_0^1 \dr\mathcal{T}\,
\mathcal{T}(y-P_z)^\beta
\Pi_\beta\big((1-\mathcal{T})(y-P_z)+P_z,p_i\big)\,,
\end{equation}
\begin{equation}\label{R}
\mathscr{R}(y,p_i):=\int \mathscr{D}\rho \int_0^1 \dr\mathcal{T}\,
(y-P_z)^\beta \Pi_\beta\big((1-\mathcal{T})(y-P_z)+P_z,p_i\big)\,.
\end{equation}
One recognizes the integrals of \eqref{zInd} and of the second
term in \eqref{division}. Note that the defined functions differ
in measure $\mathcal{T}$ only. Note also that \eqref{Z} is $y$ --
independent by \eqref{zInd}. We want to reformulate this
difference in the form of a differential equation.

Shift symmetry of $\Pi_\beta(y,p)$ \eqref{PLT} and \eqref{Pza}
yields
\begin{multline}
\Pi_\beta\big((1-\mathcal{T})(y-P_z)+P_z,p_1+a,p_2-a,\ldots\big)=\\=e^{ia_\alpha
(t+(1-\mathcal{T})(y-P_z)+P_z)^\alpha}\Pi_\beta\big((1-\mathcal{T})(y-P_z)+P_z,p_1,p_2,\ldots\big)
\end{multline}
and hence the following is true
\begin{multline}
\int \mathscr{D}\rho\int_0^1 \dr\mathcal{T} (y-P_z)^\beta
\Pi_\beta\big((1-\mathcal{T})(y-P_z)+P_z,t,p_j|\,\rho\big)e^{ia^\alpha
(t+(1-\mathcal{T})(y-P_z)+P_z)_\alpha}=\mathscr{R}(y,p_i+\epsilon_i
a)\,.
\end{multline}
Here $\epsilon_i:=(-1)^{i+1}$. Differentiating with respect to $a$
and setting $a$ to zero we obtain
\begin{multline}\label{rem1}
\int\mathscr{D}\rho \int_0^1 \dr\mathcal{T}\, (y-P_z)^\beta
\Pi_\beta\big((1-\mathcal{T})(y-P_z)+P_zt,p_j|\,\rho\big)\big((1-\mathcal{T})(y-P_z)+P_z\big)_\alpha=\\=-\left(t_\alpha+i
\sum_{j}\epsilon_j\frac{\partial}{\partial
p_j^\alpha}\right)\mathscr{R}(y,p)\,.
\end{multline}
For brevity we denote the operator on the r.h.s. as $\Delta$, i.e.
\begin{equation}
\Delta_\alpha:=\left(t_\alpha+i\sum_{j}\epsilon_j\frac{\partial}{\partial
p_j^\alpha}\right)\,.
\end{equation}
Straightforward computation yields
\begin{multline}\label{rem2}
\frac{\partial}{\partial y^\alpha} \int\mathscr{D}\rho\int_0^1 \dr\mathcal{T}\big((1-\mathcal{T})(y-P_z)+P_z\big)^\alpha (y-P_z)^\beta \Pi_\beta\big((1-\mathcal{T})(y-P_z)+P_z\big)=\\
=\left(1+y^\alpha \frac{\partial}{\partial
y^\alpha}\right)\mathscr{R}(y,p_j)-\mathscr{Z}(p_j)\,.
\end{multline}
Combining \eqref{rem1} and \eqref{rem2} together we obtain a
differential equation that relates \eqref{R} and \eqref{Z}
\begin{equation}\label{deq}
\left(1+(y+\Delta)^\alpha\frac{\partial}{\partial
y^\alpha}\right)\mathscr{R}(y,p)=\mathscr{Z}(p).
\end{equation}
General solution of this equation reads
\begin{equation}
\mathscr{R}(y,p)=\mathscr{R}_0(y,p)+\mathscr{Z}(p)\,,
\end{equation}
where $\mathscr{R}_0$ is the solution of the homogeneous equation
\begin{equation}\label{hom}
\left(1+(y+\Delta)^\alpha\frac{\partial}{\partial
y^\alpha}\right)\mathscr{R}_0(y,p)=0\,,
\end{equation}
while $\mathscr{Z}$ is any particular one of the inhomogeneous
equation. To proceed we introduce new variables
\begin{equation}\label{s}
s_\alpha:=\frac{1}{N}\sum_{j=1}^N \epsilon_jp_{j\alpha}
\end{equation}
and
\begin{equation}
\tilde{p}_k:=A_k {}^j p_j\,,\qquad A_k {}^j \epsilon_j=0\,,
\end{equation}
where $A_{k}{}^{j}$ are some constants that separate variable $s$
given by the specific linear combination of $p_i$ \eqref{s} from
the rest of $p$'s. Eq. \eqref{hom} takes the following form in
these new variables
\begin{equation}\label{hom1}
\left(1+\left(y+t+i\frac{\partial}{\partial s}\right)^\alpha
\frac{\partial}{\partial y^\alpha}
\right)\mathscr{R}_0(y,s,\tilde{p})=0\,.
\end{equation}
The left hand side operator obviously commutes with
$\frac{\partial}{\partial s}$ and hence one can seek for the
solutions of \eqref{hom1} within the eigenstates of
$\frac{\partial}{\partial s}$, i.e.
\begin{equation}\label{ES}
i\frac{\partial}{\partial
s^\alpha}\mathscr{R}_0^\sigma(y,s,\tilde{p})=\sigma_\alpha
\mathscr{R}_0^\sigma(y,s,\tilde{p})\,.
\end{equation}
For the kernel of $\frac{\partial}{\partial s}$, equation
\eqref{hom1} reads
\begin{equation}
\left(1+(y+t)^\alpha \frac{\partial}{\partial
y^\alpha}\right)\mathsf{K}(y,\tilde{p},t)=0\,.
\end{equation}
Solutions of this equation are either nonanalytic
\begin{equation}
\mathsf{K}^{reg}(y,\tilde{p},t)=\frac{1}{y^1+t^1}\Psi\left(\frac{y^1+t^1}{y^2+t^2},\tilde{p},t\right)
\end{equation}
or distributions that manifestly break Lorentz invariance
\begin{equation}
\mathsf{K}^{sng}(y,\tilde{p},t)=\psi_1(\tilde{p},t)\delta(y^1+t^1)+\psi_2(\tilde{p},t)\delta(y^2+t^2)\,,
\end{equation}
where $\psi_{1,2}$ are arbitrary functions. None of the obtained
solutions can be written as \eqref{R} in terms of well-defined
analytic $\Pi_\al$. For various eigenstates \eqref{ES} equation
\eqref{hom1} acquires the form
\begin{equation}\label{hom2}
(y+t+\sigma)^\alpha\frac{\partial}{\partial
y^\alpha}\mathscr{R}^\sigma_0(y,s,\tilde{p},t)=-\mathscr{R}^\sigma_0(y,s,\tilde{p},t)
\end{equation}
while the solution to this equation reads
\begin{equation}
\mathscr{R}^\sigma_0(y,s,\tilde{p},t)=\frac{1}{y^1+t^1+\sigma^1}\mathsf{H}\left(\frac{y^1+t^1+\sigma^1}{y^2+t^2+\sigma^2},\sigma,\tilde{p},t\right)\,
e^{i\sigma^\alpha s_\alpha}\,,
\end{equation}
where $\mathsf{H}$ is an arbitrary function. The solution to
\eqref{hom} then can be written down as an integral over all
eigenvalues
\begin{equation}
\mathscr{R}_0(y,s,\tilde{p},t)=\int \dr^2\sigma
\frac{1}{y^1+t^1+\sigma^1}\mathsf{H}\left(\frac{y^1+t^1+\sigma^1}{y^2+t^2+\sigma^2},\sigma,\tilde{p},t\right)
e^{i\sigma^\alpha s_\alpha}.
\end{equation}
This expression is nonanalytic in $y$ and thus it cannot be a
suitable candidate for \eqref{R}. Once again one can consider
solutions that contain distributions, i.e.
\begin{equation}
\mathscr{R}_0 (y,s,\tilde{p},t)=\int
\dr^2\gs\left(\delta(y^1+\sigma^1+t^1)\mathsf{h}_1(\sigma,\tilde{p},t)e^{i\sigma^\alpha
s_\alpha}+\delta(y^2+\sigma^2+t^2)\mathsf{h}_2(\sigma,\tilde{p},t)e^{i\sigma^\alpha
s_\alpha}\right)\,.
\end{equation}
Integrating out delta-functions one arrives at
\begin{align}
\mathscr{R}_0 (y,s,\tilde{p},t)=&\int \dr\gs^2
\mathsf{h}_1(-y^1-t^1,
\gs^2,\tilde{p},t)e^{-i(y^1+t^1)s_1+i\gs^2s_2}+\\
&+\int \dr\gs^1 \mathsf{h}_2(\gs^1, -y^2-t^2, \tilde p,
t)e^{i\gs^1 s_1-i(y^2+t^2)s_2}\,.
\end{align}
While the latter solution can be perfectly analytic with respect
to variables $y, p$ and $t$, it manifestly breaks Lorentz symmetry
and for that reason should be excluded. Our analysis leads to the
conclusion that $\mathscr{R}$ coincides with $\mathscr{Z}$, which
is $y$-independent
\be\label{Z=R}
\mathscr{R}=\mathscr{Z}(p)\,.
\ee
Recall, the latter equation holds provided $z$-independence
\eqref{zInd}, shift symmetry \eqref{PLT} and Lorentz invariance
imposed. Hence the remainder after being integrated over $\rho$ in
the division formula \eqref{division} is equal to zero, i.e.
\begin{equation}\label{remZ}
\frac{\partial}{\partial y^\alpha}\int \mathscr{D}\rho \int_0^1
\dr\mathcal{T}\, (y-P_z)^\sigma
\Pi_\sigma\big((1-\mathcal{T})(y-P_z)+P_z,t,p_j|\,\rho\big)=\ff{\p}{\p
y^\al}\mathscr{Z}(p)\equiv0\,.
\end{equation}

\section{Locality and field redefinitions}\label{Locality}
To analyze locality we start with identity \eqref{division1}. The
result of the previous section implies that after integration over
$\rho$ and by virtue of \eqref{remZ} the division formula turns
into
\begin{equation}\label{noRem}
\int \mathscr{D}\rho\,
\Pi_{\alpha}\big(y,t,p_j|\,\rho\big)=\int\mathscr{D}\rho\,
(y-\underline{P_z})_\alpha \int_0^1
\dr\mathcal{T}\frac{\partial}{\partial y^\beta}
\Pi^\beta\big((1-\mathcal{T})(y-P_z)+P_z,t,p_j|\,\rho\big)\,.
\end{equation}
Recall that function $\Pi_{\al}$ is free from nonlocal
$pp$-contractions. Despite the fact that the l.h.s. is manifestly
local it does not immediately imply that the corresponding vertex
\eqref{Vd} is local. This is due to the presence of
$\rho$-dependent $P_z$ underlined above. As a matter of principle
the two terms on the right can be nonlocal separately still
leading to local l.h.s. of \eqref{noRem}. Among these two below
\begin{multline}\label{two}
y_\alpha\int\mathscr{D}\rho  \int_0^1
\dr\mathcal{T}\frac{\partial}{\partial y^\beta}
\Pi^\beta\big((1-\mathcal{T})(y-P_z)+P_z,t,p_j|\,\rho\big)-\\-\int\mathscr{D}\rho\,
P_{z\alpha}  \int_0^1 \dr\mathcal{T}\frac{\partial}{\partial
y^\beta}
\Pi^\beta\big((1-\mathcal{T})(y-P_z)+P_z,t,p_j|\,\rho\big)
\end{multline}
the first one is simply the vertex times $y_\alpha$ (see section
\ref{divS}). Our goal is to show that it is local by virtue of the
locality of the second one. Contracting both sides of
\eqref{noRem} with $y_\alpha$ one obtains
\begin{equation}\label{YC}
\int \mathscr{D}\rho\, y^\alpha
\Pi_{\alpha}\big(y,t,p_j|\,\rho\big)=-\int\mathscr{D}\rho\,
y^\alpha P_{z\alpha} \int_0^1
\dr\mathcal{T}\frac{\partial}{\partial y^\beta}
\Pi^\beta\big((1-\mathcal{T})(y-P_z)+P_z,t,p_j|\,\rho\big)\,.
\end{equation}
Since the l.h.s is local so is the right one. The latter
expression therefore means that any non-locality that might be
present in the second term of \eqref{two} is proportional to
$y_{\al}$.

Now we expand the second term of \eqref{two} extracting the piece
proportional to $y_{\al}$.
\begin{multline}\label{Loc1}
\int \mathscr{D} \rho\, P_{z\alpha}\int_0^1 \dr\mathcal{T}\, (1-\mathcal{T}) \partial_\beta \Pi^\beta \big((1-\mathcal{T})(y-P_z)+P_z,t,p_j|\,\rho\big)=\\=\int\mathscr{D}\rho \, P_{z\alpha} \int_0^1 \dr\mathcal{T}\, (1-\mathcal{T})\partial_\beta\Pi^\beta(\mathcal{T}P_z,t,p_j|\,\rho)+\\
+\sum_{n=1}^\infty\int \mathscr{D}\rho \, P_{z\alpha}\int_0^1
\dr\mathcal{T}\,
\frac{(1-\mathcal{T})^{n+1}}{n!}y^{\gamma_1}\ldots y^{\gamma_n}
\partial_{\gamma_1}\ldots
\partial_{\gamma_n}\partial_\beta\Pi^\beta(\mathcal{T}P_z,t,p_j|\,\rho)\,.
\end{multline}
Here $\partial_{\beta}$ is the derivative of $\Pi^\alpha$ with
respect to the full first argument. Applying the Schouten identity
\begin{equation}
P_{z\alpha}y^\sigma \partial_{\sigma}\equiv y_\alpha P_z {}^\sigma
\partial_{\sigma}+y^\sigma P_{z\sigma} \partial_\alpha
\end{equation}
one can rewrite \eqref{Loc1} as
\begin{multline}\label{Loc2}
\int \mathscr{D} \rho\, P_{z\alpha}\int_0^1 \dr\mathcal{T}\, (1-\mathcal{T}) \partial_\beta \Pi^\beta \big((1-\mathcal{T})(y-P_z)+P_z,t,p_j|\,\rho\big)=\\=\int\mathscr{D}\rho \, P_{z\alpha} \int_0^1 \dr\mathcal{T}\, (1-\mathcal{T})\partial_\beta\Pi^\beta(\mathcal{T}P_z,t,p_j|\,\rho)+\\
+y_\alpha\sum_{n=1}^\infty\int \mathscr{D}\rho \int_0^1 \dr\mathcal{T}\, \frac{(1-\mathcal{T})^{n+1}}{n!}y^{\gamma_2}\ldots y^{\gamma_n}\frac{\partial}{\partial \mathcal{T}} \partial_{\gamma_2}\ldots \partial_{\gamma_n}\partial_\beta\Pi^\beta(\mathcal{T}P_z,t,p_j|\,\rho)+\\
+\sum_{n=1}^\infty\int \mathscr{D}\rho  \big(y^\sigma
P_{z\sigma}\big)\int_0^1 \dr\mathcal{T}\,
\frac{(1-\mathcal{T})^{n+1}}{n!}y^{\gamma_2}\ldots y^{\gamma_n}
\partial_{\gamma_2}\ldots \partial_{\gamma_n}\partial_\alpha
\partial_\beta\Pi^\beta(\mathcal{T}P_z,t,p_j|\,\rho)\,.
\end{multline}
Performing partial integration with respect to $\mathcal{T}$ one
obtains \eqref{Loc2} in the form
\begin{multline}\label{Loc3}
\int \mathscr{D} \rho\, P_{z\alpha}\int_0^1 \dr\mathcal{T}\,
(1-\mathcal{T}) \partial_\beta \Pi^\beta
\big((1-\mathcal{T})(y-P_z)+P_z,t,p_j|\,\rho\big)=\\=\int\mathscr{D}\rho
\, P_{z\alpha} \int_0^1 \dr\mathcal{T}\,
(1-\mathcal{T})\partial_\beta\Pi^\beta(\mathcal{T}P_z,t,p_j|\,\rho)-\\
-y_\alpha\sum_{n=1}^\infty \frac{1}{n!}\int \mathscr{D}\rho  \, y^{\gamma_2}\ldots y^{\gamma_n}\partial_{\gamma_2}\ldots \partial_{\gamma_n} \partial_\beta\Pi^\beta(\mathcal{T}P_z,t,p_j|\,\rho)\Big|_{\mathcal{T}=0}+\\
+y_\alpha \sum_{n=1}^\infty\int\mathscr{D}\rho \int_0^1 \dr\mathcal{T}\, \frac{(1-\mathcal{T})^{n}(n+1)}{n!}y^{\gamma_2}\ldots y^{\gamma_n} \partial_{\gamma_2}\ldots \partial_{\gamma_n}\partial_\beta\Pi^\beta(\mathcal{T}P_z,t,p_j|\,\rho)+\\
+\sum_{n=1}^\infty\int \mathscr{D}\rho  \big(y^\sigma
P_{z\sigma}\big)\int_0^1 \dr\mathcal{T}\,
\frac{(1-\mathcal{T})^{n+1}}{n!}y^{\gamma_2}\ldots y^{\gamma_n}
\partial_{\gamma_2}\ldots \partial_{\gamma_n}\partial_\alpha
\partial_\beta\Pi^\beta(\mathcal{T}P_z,t,p_j|\,\rho)\,.
\end{multline}
Introducing an additional integration variable
\begin{equation}
\frac{1}{n!}=\int_0^1 \dr\xi \frac{\xi^{n-1}}{(n-1)!}\,.
\end{equation}
one is able to obtain \eqref{Loc3} in the following form
\begin{multline}\label{Loc4}
\int \mathscr{D} \rho\, P_{z\alpha}\int_0^1 \dr\mathcal{T}\, (1-\mathcal{T}) \partial_\beta \Pi^\beta \big((1-\mathcal{T})(y-P_z)+P_z,t,p_j|\,\rho\big)=\\
=\int\mathscr{D}\rho \, P_{z\alpha} \int_0^1 \dr\mathcal{T}\,
(1-\mathcal{T})\partial_\beta\Pi^\beta(\mathcal{T}P_z,t,p_j|\,\rho)
-y_\alpha \int \mathscr{D}\rho\int_0^1 \dr\xi\, \partial_\beta\Pi^\beta(\xi y,t,p_j|\,\rho)+\\
+y_\alpha \int \mathscr{D}\rho \int_0^1 \dr\mathcal{T}\, (1-\mathcal{T}) \partial_\beta \Pi^\beta\big((1-\mathcal{T})y+\mathcal{T}P_z,t,p_j|\,\rho\big)+\\
+y_\alpha \int\mathscr{D}\rho \int_0^1 \dr\xi\int_0^1 \dr\mathcal{T}\, (1-\mathcal{T}) \partial_\beta\Pi^\beta\big((1-\mathcal{T})\xi y+\mathcal{T}P_z,t,p_j|\,\rho\big)+\\
+\int\mathscr{D}\rho \int_0^1 \dr\xi \int_0^1 \dr\mathcal{T}\,
(1-\mathcal{T})^2 \big(y^\sigma
P_{z\sigma}\big)\partial_\alpha\partial_\beta\Pi^\beta\big((1-\mathcal{T})\xi
y+\mathcal{T}P_z,t,p_j|\,\rho\big)\,.
\end{multline}
Plugging the obtained result into \eqref{noRem}  we obtain
\begin{multline}\label{Loc5}
\int \mathscr{D}\rho\,
\Pi_{\alpha}(y,t,p_j|\,\rho)=-\int\mathscr{D}\rho \, P_{z\alpha}
\int_0^1 \dr\mathcal{T}\,
(1-\mathcal{T})\partial_\beta\Pi^\beta(\mathcal{T}P_z,t,p_j|\,\rho)
+\\+y_\alpha \int \mathscr{D}\rho\int_0^1 \dr\xi\, \partial_\beta\Pi^\beta(\xi y,t,p_j|\,\rho)-\\
-y_\alpha \int\mathscr{D}\rho \int_0^1 \dr\xi\int_0^1 \dr\mathcal{T}\, (1-\mathcal{T}) \partial_\beta\Pi^\beta\big((1-\mathcal{T})\xi y+\mathcal{T}P_z,t,p_j|\,\rho\big)-\\
-\int\mathscr{D}\rho \int_0^1 \dr\xi \int_0^1 \dr\mathcal{T}\,
(1-\mathcal{T})^2 \big(y^\sigma
P_{z\sigma}\big)\partial_\alpha\partial_\beta\Pi^\beta\big((1-\mathcal{T})\xi
y+\mathcal{T}P_z,t,p_j|\,\rho\big)\,.
\end{multline}
Here terms that are not proportional to $y_\alpha$ are local (at
least being combined together, see \eqref{YC}). There is also the
manifestly local term proportional to $y_\alpha$ in the first
line, hence the remaining $y_\alpha$-proportional term is local as
a consequence. The leftover term, namely
\begin{equation}
\widetilde{\Upsilon}(y,t,p_j)=\int\mathscr{D}\rho \int_0^1
\dr\xi\int_0^1 \dr\mathcal{T}\, (1-\mathcal{T})
\partial_\beta\Pi^\beta\big((1-\mathcal{T})\xi
y+\mathcal{T}P_z,t,p_j|\,\rho\big)
\end{equation}
is not proportional to the vertex (see \eqref{div4} and below)
\begin{equation}
\Upsilon(y,t,p_j)=\int \mathscr{D}\rho \int_0^1 \dr\mathcal{T}\,
(1-\mathcal{T})\partial_\beta\Pi^\beta\big((1-\mathcal{T})y+\mathcal{T}P_z,t,p_j|\,\rho\big)\,.
\end{equation}
However the exact relation is given by
\begin{equation}
\widetilde{\Upsilon}(y,t,p_j)=\int_0^1 \dr\xi \, \Upsilon(\xi
y,t,p_j)\,.
\end{equation}
This relation in particular implies that $\Upsilon$ is local since
the new integration over $\xi$ does not involve derivatives
$\{p_i\}$. Moreover one can invert this integral over $\xi$, i.e.
\begin{equation}
\Upsilon(y,t,p_j)=\left(y^\alpha \frac{\partial}{\partial
y^\alpha}+1\right)\widetilde{\Upsilon}(y,t,p_j)\,.
\end{equation}
which implies that vertex $\Upsilon$ is local.

\subsection*{Field redefinitions} HS equations
\eqref{1form}-\eqref{0form} possess natural ambiguity in field
redefinitions that can be written as follows
\begin{align}
&C\; \rightarrow\; C+\sum_n
b_n(y,p_1,\ldots,p_n)\underbrace{C\ldots C}_n\,,\label{Ctr}\\
&\go\; \rightarrow\; \go+\sum_n w_n(y,p_1,\ldots,p_n,
t)\go\underbrace{C\ldots C}_n\,,\label{wtr}
\end{align}
where $b_n$ and $w_n$ are some arbitrary functions. If these
functions are (non)-polynomial in various contractions $pp$ then
the corresponding field redefinitions are (non)-local. Suppose one
has a local vertex $\Upsilon$. By applying non-local field
redefinition one arrives at non-local $\Upsilon$. Interestingly,
however, local redefinitions do not necessarily preserve locality
of the vertex \cite{Vasiliev:2022med}. This happens because HS
spectrum is unbounded from above. For example, vertex
$\Upsilon_{\go\go C}$ being linear in $C$ is always local, but it
has to be be strictly {\it ultra-local}\footnote{See
\cite{Didenko:2018fgx} for the definition of ultra-locality.} in
order to be consistent with locality of $\Upsilon_{\go CC}$.
Nevertheless, the field redefinition that relates local form of
$\Upsilon_{\go\go C}$ to its ultra-local one is local. This
example (see \cite{Didenko:2022qga} for more detail) illustrates
that the locality of field redefinitions alone is not enough for
having HS interactions local.

We would like to argue that locality of HS vertex is still
preserved for spin-local $b_n$ and spin ultra-local $w_n$ provided
these functions in addition are properly shift-symmetric. Namely,
\begin{align}
&b_n(y,p_1+a,p_2-a,\ldots, p_n+(-)^{n+1}a)=e^{ia_\alpha
y^\alpha}b_n(y,p_1,p_2,\ldots,p_n)\,,\label{PLTredef}\\
&w_n(y-a,p_1+a,p_2-a,\ldots, p_n+(-)^{n+1}a, t)=e^{ia_\alpha
t^\alpha}w_n(y,p_1,p_2,\ldots,p_n, t)\label{PLTredef1}
\end{align}
should respect HS locality under \eqref{Ctr} and \eqref{wtr}.
Field redefinitions that satisfy
\eqref{PLTredef}-\eqref{PLTredef1} are called {\it
shift-symmetric}. Supporting evidence in favor of this conjecture
comes from the class preserving theorems proven in \cite{Gelfond:2018vmi},
where it was shown that star products and proper homotopy shifts
respect certain structures that underlie the observed shift
symmetry.

An example of field redefinition \eqref{PLTredef} is in order. The
local quadratic $C^2$ vertex that carries minimum number of
derivatives originally found in \cite{Vasiliev:2017cae} differs from the one
obtained later via shifted homotopies in \cite{Didenko:2018fgx} by
\begin{equation}
C\; \rightarrow\; C+\frac{\eta}{2}\int_0^1\dr t\,
C(ty)C((t-1)y)=\frac{\eta}{2}\int_0^1 \dr t\,
e^{iy^\alpha(tp_1+(t-1)p_2)_\alpha} CC\,,
\end{equation}
which obviously enjoys \eqref{PLTredef} and leads to a local
contribution at $C^3$.

While we do not have proof for the claim
\eqref{PLTredef}-\eqref{PLTredef1} beyond order $C^3$ within the
Vasiliev generating equations, it has recently received a strong
all order support from \cite{Didenko:2022qga}.

\section{Conslusions and open questions}
In this paper we specify and prove the $z$ -- dominance conjecture
of \cite{Gelfond:2018vmi,Gelfond:2019tac}. Originally formulated
as a statement that the $z$-dominated terms arising in the process
of solving Vasiliev equations can not appear in the
$z$-independent HS vertices it is relaxed here by assuming their
appearance in no more than a polynomial way. This modification
still agrees with the assertion of \cite{Gelfond:2018vmi}. Namely,
the apparent $z$-dominated non-localities resulted from the
Vasiliev master fields leave HS vertices spin-local. However, we
have not been able to prove this statement as is and have further
specified it by adding a symmetry assumption. The fact that
$z$-dominated terms cut short polynomially in the final vertices
as observed in \cite{Gelfond:2021two} seems to rely on a certain
symmetry. The so-called shift symmetry has been indeed identified
at first few orders based on explicit results of
\cite{Didenko:2018fgx,Didenko:2020bxd,Didenko:2019xzz}. While we
do not have proof of this symmetry beyond order $C^3$ within the
Vasiliev equations, we assume it is still there at higher orders.
Its origin can be traced back to the locality theorems of
\cite{Gelfond:2018vmi}. Moreover, recently in
\cite{Didenko:2022qga} the shift symmetry has been shown to
manifest to all orders within a different set of generating
equations.

The main message of our investigation is the following. If in
solving of the Vasiliev system for a HS vertex one encounters
contributions which (i) carry $z$-dominated non-localities, (ii)
are shift symmetric (iii) are Lorentz invariant, -- then the
corresponding vertex is spin-local. Despite a diligent effort, we
were not able to relax (ii) which plays a crucial role in our
proof and entails (iii).

The proof is based on the following technical steps
\begin{enumerate}
\item  We bring HS vertex to the form \eqref{Start} modulo local
$\dr_z$ -- cohomologies using operator $\ord_{-\infty}$. This
representation has a somewhat factorized form with all the vertex
details stored in a local (contains finitely many
$pp$-contractions) function $\Pi_{\al}(y, t, p_i|\,\rho)$ and
spinor $P_{z\al}(t, p_i|\,\rho)$.

\item Examine the $z$-independence requirement of \eqref{Start}
which gives rise to \eqref{zInd1}. The latter equation contains
infinitely many integration constraints associated with each power
of $z$. We then observe that all these get manifestly resolved if
$\Pi_\al$ has a form \eqref{div} that introduces the vertex
function $\Pi$.

\item The previous step suggests to decompose $\Pi_{\al}$ into two
pieces. The one that manifestly resolves $z$-independence
condition \eqref{zInd1} via the vertex function $\Pi$ and the
remainder. The decomposition is achieved via division formula
\eqref{division1}. Surprisingly, up to an integration measure the
remainder looks precisely as $z$-independence condition
\eqref{zInd1}. Had it been literally the same the proof would have
stopped here.

\item The final step is to show that the $z$-independence still
implies the vanishing of the remainder. In that case locality of
the resulting vertex is guaranteed while the $z$ -- dominance
conjecture is justified. This step in fact is most involved and
seems to require an additional shift symmetry assumption
\eqref{PLT}-\eqref{PLT1}, which allows us to arrive at certain
differential equation \eqref{deq} that relates the remainder with
the $z$-independence condition. By analyzing its solution space
and restricting it to analytic and Lorentz invariant functions one
proves that this space is empty thus implying that the remainder
does vanish.
\end{enumerate}
Interestingly, a particular role of the shift symmetry is in
making $z$-independence condition \eqref{zInd1} satisfied for all
$z$ provided it is fulfilled at $z=0$. It does not seem to work
other way around though, i.e. an infinite number of conditions
falling off from \eqref{zInd1} for each power of $z$ does not
bring the shift symmetry by default.

An appeal to the shift symmetry may look unsatisfactory since a
chance that infinite chain of the $z$-independence constraints in
\eqref{zInd1} might yield the remainder zero under a milder extra
assumption is not excluded. Still, we tend to think of this
symmetry as of something deeply interrelated to the locality much
as the homotopy shifts from \cite{Gelfond:2018vmi} underlying the locality
theorems are. That being said the all order local holomorphic HS
vertices recently analyzed in \cite{Didenko:2022qga} are all shift-symmetric. An
interesting feature of the shift symmetry that can be seen at
orders $C^2$ and $C^3$ is it does not manifest itself in a sense
of having particular simple transformations under $p_i\to
p_i+(-)^{i+1}a$ within a given individual vertex contribution of
the Vasiliev system unless the $z$-dominated terms dropped off.
Nevertheless, the sum of all of them leads to a shift symmetric
vertex, e.g. \eqref{3O}, \eqref{2O}. In this respect the basic
objects the symmetry naturally acts upon are $\Pi_{\al}$ and $P_z$
from \eqref{Start}. These show up in each contribution.

Another point that our analysis suggests concerns the problem of
the admissible field redefinitions. It is known (see e.g.
\cite{Didenko:2022qga} for an illustrative discussion) that local
field redefinitions do not necessarily preserve locality of HS
vertices. We argue that local field redefinitions
\eqref{Ctr}-\eqref{wtr} keep HS vertex spin-local provided they
are shift-symmetric \eqref{PLTredef}-\eqref{PLTredef1}.

The motivation of the proposed investigation partly rests on the
desire to have a qualitative answer whether the Vasiliev
generating procedure gives rise to local HS vertices\footnote{For
a recent interesting view on HS locality see also
\cite{Lysov:2022nsv}.} without their explicit calculation which
can be really a formidable one in practice \cite{Gelfond:2021two}.

The procedure of obtaining vertices in \cite{Gelfond:2021two}
roughly can be described as follows. Starting with the sum of
expressions of the kind \eqref{generic} one singles out all the
cohomology terms by virtue of partial integration over
$\mathcal{T}$. The remaining expression can be zero or contain
more cohomologies. If it is not zero then to proceed one rewrites
measure $R(\rho)$ in \eqref{toUniform} as a derivative of some
$\rho$'s and integrates by parts. Then once again one singles out
all the cohomologies and investigates if the remaining expression
is zero. If not, one continues with partial integration over
various $\rho$'s. These partial integrations over $\rho$ bring
more and more derivatives $p_j$ in the pre-exponential at each new
step. Thus being polynomial in the beginning it may turn into
power series if the process just described does not terminate
after a finite number of steps. This indicates that the $z$ --
dominance lemma as formulated in
\cite{Gelfond:2018vmi,Gelfond:2019tac} and proven at $C^2$ may not
necessarily imply spin locality at higher orders without extra
assumptions considered in our paper.

In conclusion let us stress that even though the conditions for
the $z$ -- dominance conjecture are verified up to the third order
vertices within the Vasiliev system, our analysis is not confined
to a specific number of $C$ fields entering vertex. Thus if
someone is able to provide the reduced dominated contributions
$f_k(z,y,t,p_j)$ for a vertex of, say, order $n$ which enjoy
\begin{equation}
f_k(z,y,t,p_1+a,p_2-a,\ldots ,p_n-(-1)^n
a)=e^{ia_\alpha(t+y)^\alpha}f_k(z,y,t,p_1,p_2,\ldots ,p_n)
\end{equation}
then the corresponding vertex is spin local. Unfortunately we are
still unable to provide a simple recipe for obtaining manifestly
spin-local vertices from the $z$-dominated expressions.

Another open question is the origin of the shift symmetry beyond
order $C^3$ within the Vasiliev theory. It plays significant role
in our analysis. We expect this problem can be analyzed at the
level of operations on the Vasiliev equations along the lines of
\cite{Gelfond:2018vmi}. The way it was proven to all orders in
\cite{Didenko:2022qga} is based on the analysis of structures that
star product and standard contracting homotopy bring within a
prescribed class of functions.

\section*{Acknowledgment}
We would like to thank Sasha Smirnov for fruitful discussions on
several technical issues of our analysis. In particular, the
speculation on analytic but not Lorentz invariant solutions of
\eqref{deq} was very stimulating. The authors are also grateful to
O.A. Gelfond and M.A. Vasiliev for many valuable comments on the
draft. This research was supported by the Russian Science
Foundation grant 18-12-00507.

\newcounter{appendix}
\setcounter{appendix}{1}
\renewcommand{\theequation}{\Alph{appendix}.\arabic{equation}}
\addtocounter{section}{1} \setcounter{equation}{0}
\renewcommand{\thesection}{\Alph{appendix}.}
%\newpage

\addcontentsline{toc}{section}{Appendix A. Second order
contributions to vertex $\Upsilon^\eta_{\omega CC}$}

\section*{Appendix A. Second order contributions to vertex $\Upsilon^\eta_{\omega CC}$}\label{A}

Here we provide an example of explicit contributions that come out
of the Vasiliev generating system to form vertex $\Upsilon_{\omega
CC}^\eta$. Up to this order eq. \eqref{ver0} reads
\begin{equation}\label{CCver}
\Upsilon_{\omega CC}^\eta=-\dr_x B_2^\eta\Big|_{\go CC}-\omega
\ast B_2^\eta-W_{1\, \go C}^\eta \ast C\,,
\end{equation}
which arranges into
\be
\Upsilon_{\go
CC}=\ff{\eta}{2}\int_{0}^{1}\dr\rho\,\Pi^{\al}(y;p_1,p_2,t|\,\rho)\circledast\int_{0}^{1}
\dr\mathcal{T} (1-\mathcal{T})z_{\al} e^{i\mathcal{T}
z_{\al}(y+p_1+p_2+t)^{\al}}\go CC\,,
\ee
where $\Pi_{\al}$ collects three different contributions from
\eqref{CCver}
\be
\Pi_{\al}=\Pi_{\al}\Big|_{\dr_x
B_2}+\Pi_{\al}\Big|_{\go*B_2}+\Pi_{\al}\Big|_{W_1*C}
\ee
with each individual part being

\begin{align}
&\Pi_{\al}\Big|_{\dr_x B_2}=y_{\al} e^{i((1-\rho)p_2-\rho
(p_1+t))^{\gb}y_{\gb}-ip_1^{\gb}t_{\gb}}\,,\label{pi1}\\
&\Pi_{\al}\Big|_{\go*B_2}=-(y+t)_{\al} e^{i((1-\rho)p_2-\rho
p_1)^{\gb}(y+t)_{\gb}-it^{\gb}y_{\gb}}\,,\label{pi2}\\
&\Pi_{\al}\Big|_{W_1*C}=t_{\al}e^{i(p_2-\rho
t)^{\gb}y_{\gb}}\,.\label{pi3}
\end{align}
Note that for all of them there is one and the same $P_z$ which is
$\rho$-independent
\be
P_{z}=-p_1-p_2-t\,.
\ee
$Z$-independence condition \eqref{2nd} in this case amounts to
\be
(y+p_1+p_2+t)^{\al}\int_{0}^{1}\dr\rho\,\Pi_{\al}=0\,,
\ee
which is resolved in terms of the $\rho$-integral of the vertex
function $\Pi$
\be
\int_{0}^{1}\dr\rho\,\Pi_{\al}:=(y+p_1+p_2+t)_{\al}\int_{0}^{1}\dr\rho\,
\Pi=(y+p_1+p_2+t)_{\al}\Phi\,.
\ee
The latter can be conveniently found using \eqref{2ndver}
\be\label{api}
\Phi=\ff{t^{\al}\int_{0}^{1}\dr\rho\,\Pi_{\al}}{t^{\gb}(y+p_1+p_2)_{\gb}}\,.
\ee
By introducing the following auxiliary function
\be\label{vfunc}
\Psi(y; p_1, p_2, t)=\ff{y\cdot t}{t\cdot
(y+p_1+p_2)}\int_{0}^{1}\dr\rho\, e^{i((1-\rho)p_2-\rho
p_1)^{\al}y_{\al}}
\ee
and using \eqref{pi1}-\eqref{pi3} one rewrites \eqref{api} as
\be
\Phi(y;p_1,p_2,t)=\Psi(y; p_1+t, p_2,
t)e^{-ip_{1}^{\al}t_{\al}}-\Psi(y+t; p_1,p_2,
t)e^{-it^{\al}y_{\al}}\,.
\ee
Non-analyticity in \eqref{vfunc} manifestly cancels out in $\Phi$
if rewritten via an extra integral over $\rho'$
\be
\Phi=\int_{[0,1]^2} \dr\rho\, \dr\rho'\, (1-\rho)y\cdot t\,
e^{i\rho'[((1-\rho)p_2-\gs(p_1+t))\cdot y-p_1\cdot
t]+i(1-\rho')[((1-\rho)p_2-\rho p_1)\cdot(y+t)-t\cdot y]}
\ee
and so the vertex is
\be
\Upsilon_{\go CC}:=\ff{i\eta}{2}\Phi(y; p_1, p_2, t)\,\go CC\,.
\ee

\renewcommand{\theequation}{\Alph{appendix}.\arabic{equation}}
\addtocounter{appendix}{1} \setcounter{equation}{0}
\addtocounter{section}{1} \addcontentsline{toc}{section}{Appendix
B. Third order contributions to vertex
$\Upsilon^{\eta\eta}_{CCC\go}$}

\section*{Appendix B. Third order contributions to vertex $\Upsilon^{\eta\eta}_{CCC\go}$}\label{B}

Here we consider an example of $C^3$-vertex from the holomorphic
0-form sector of the Vasiliev equations. From the point of view of
representation \eqref{result} its contributions reveal some new
features. Namely, as opposed to the case of $C^2$-vertices, $P_z$
becomes $\rho$-dependent. While this leads to a formidable
complication in obtaining manifest $z$-independent expressions of
the HS vertex (see \cite{Gelfond:2021two}), it causes no trouble
in checking conditions for $z$ -- dominance conjecture and
therefore in the status of locality. The expressions below are
just for illustration.

The vertex contains five individual contributions
\begin{equation}\label{verCCC}
\Upsilon_{CCC\go}^{\eta\eta}=-\dr_x
B_2^\eta\Big|_{CCC\go}+B_2^\eta\ast W_{1\, \go C}^\eta-\dr_x
{B}_3^{\eta\eta}\Big|_{CCC\go}+B_3*\go+C\ast{W}^{\eta\eta}_{2\,
CC\go}\,,
\end{equation}
where we stick to the particular $CCC\go$ -- ordering in this one.
Each individual contribution has the form \eqref{generic}. These
expressions in their explicit form are provided below
\begin{multline}\label{START}
\dr_x {B}_3^{\eta\eta}\big|_{ CCC\go}\approx\frac{i\eta^2}{4} \int_0^1 \dr\mathcal{T}\, \mathcal{T}  \int \dr^3 \rho_+ \delta\left(1-\sum_{i=1}^3 \rho_i\right)   \int_0^1 \dr\xi\, \frac{\rho_1\, (z_\alpha y^\alpha)^2 e^{i\mathcal{T}\, z_\alpha y^\alpha} }{(\rho_1+\rho_2)(\rho_1+\rho_3)}\times\\
\times\exp\Big\{i\mathcal{T} z^\alpha\Big(-(\rho_1+\rho_3)p_{1\alpha}+(\rho_2-\rho_3)p_{2\alpha}+(\rho_1+\rho_2)(t_{\alpha}+p_{3\alpha})\Big)-ip_{3 \alpha}t^\alpha\\
+i(1-\xi)y^\alpha\left(\frac{\rho_1}{\rho_1+\rho_2}p_{1\alpha}-\frac{\rho_2}{\rho_1+\rho_2}p_{2\alpha}\right)+i\xi\, y^\alpha\left(\frac{\rho_1}{\rho_1+\rho_3}(t_{\alpha}+p_{3\alpha})-\frac{\rho_3}{\rho_1+\rho_3}p_{2\alpha}\right)\Big\}CCC\go\, ,
\end{multline}
\begin{multline}
 {B}_3^{\eta\eta}\ast \go \approx\frac{i\eta^2}{4} \int_0^1 \dr\mathcal{T}\, \mathcal{T}  \int \dr^3 \rho_+ \delta\left(1-\sum_{i=1}^3 \rho_i\right)   \int_0^1 \dr\xi\, \frac{\rho_1\, \left[z_\alpha\left(y^\alpha-t^\alpha\right)\right]^2 e^{i\mathcal{T}\, z_\alpha y^\alpha} }{(\rho_1+\rho_2)(\rho_1+\rho_3)}\times\\
\times\exp\Big\{i\mathcal{T} z^\alpha\Big(t_{\alpha}-(\rho_1+\rho_3)p_{1\alpha}+(\rho_2-\rho_3)p_{2\alpha}+(\rho_1+\rho_2)p_{3\alpha}\Big)+i y^\alpha t_{\alpha}\\
+i(1-\xi)y^\alpha\left(\frac{\rho_1}{\rho_1+\rho_2}p_{1\alpha}-\frac{\rho_2}{\rho_1+\rho_2}p_{2\alpha}\right)+i\xi\, y^\alpha\left(\frac{\rho_1}{\rho_1+\rho_3}p_{3\alpha}-\frac{\rho_3}{\rho_1+\rho_3}p_{2\alpha}\right)\\
-i\frac{(1-\xi)\rho_1}{\rho_1+\rho_2}p_{1 \alpha}t^\alpha+i\left(\frac{(1-\xi)\rho_2}{\rho_1+\rho_2}+\frac{\xi\rho_3}{\rho_1+\rho_3}\right) p_{2 \alpha}t^\alpha-i\frac{\xi\rho_1}{\rho_1+\rho_3}p_{3 \alpha}t^\alpha\Big\} CCC\go\,,
\end{multline}
\begin{multline}
\dr_x B_2^{\eta\, loc}\big|_{CCC\go}\approx-\frac{i\eta^2}{4}\int_0^1 \dr\mathcal{T}\int_0^1 \dr\xi\int \dr^3 \rho_+\, \delta\left(1-\sum_{i=1}^3\rho_i\right)\left(z_\alpha y^\alpha\right)\Big[i\left(\mathcal{T}z^\alpha+(1-\xi)y^\alpha\right)t_{ \alpha}\Big]\times\\
\times \exp\Big\{i\mathcal{T} z_\alpha y^\alpha -i(1-\rho_2)p_{3\alpha}t^\alpha+i\rho_2 p_{2 \alpha}t^\alpha+i\mathcal{T}z^\alpha\Big(-p_{1 \alpha}-(\rho_2+\rho_3)p_{2 \alpha}+\rho_1 p_{3\alpha}+(\rho_1+\rho_2)t_{\alpha}\Big)\\
+i y^\alpha\Big(\xi p_{1 \alpha}-(1-\xi)(\rho_2+\rho_3) p_{2 \alpha}+(1-\xi)\rho_1 p_{3\alpha}+(1-\xi)(\rho_1+\rho_2)t_{\alpha}\Big)\Big\}CCC\go\,,
\end{multline}
\begin{multline}
B_2^{\eta\, loc}\ast W_{1\, C\go}^\eta\approx\frac{i\eta^2}{4}\int_0^1 \dr\mathcal{T} \int_0^1 \dr\Sigma \int \dr^3\rho_+\, \frac{\delta\left(1-\sum_{i=1}^3 \rho_i\right)}{\rho_1+\rho_2}\Big[z_\alpha y^\alpha+\Sigma z^\alpha t_{\alpha}\Big]\left(iz^\gamma t_{\gamma}\right)\times\\
\times \exp\Big\{i\mathcal{T} z_\alpha y^\alpha-i(1-\Sigma)p_{3 \alpha}t^\alpha+i\frac{\rho_1 \Sigma}{\rho_1+\rho_2}p_{1 \alpha}t^\alpha-i\frac{\rho_2 \Sigma}{\rho_1+\rho_2}p_{2 \alpha}t^\alpha\\
+i\mathcal{T}z^\alpha\Big(-(\rho_3+\rho_1)p_{1 \alpha}-(\rho_3-\rho_2)p_{2 \alpha}+(\rho_1+\rho_2-\Sigma\rho_3)t_{\alpha}+(\rho_1+\rho_2)p_{3\alpha}\Big)\\
+iy^\alpha\Big(\frac{\rho_1}{\rho_1+\rho_2}p_{1 \alpha}-\frac{\rho_2}{\rho_1+\rho_2}p_{2\alpha}-\Sigma t_{\alpha}\Big)\Big\}CCC\go\,,
\end{multline}
\begin{multline}\label{FINISH}
C\ast {W}_{2\, CC\go}^{\eta\eta}\approx -\frac{\eta^2}{4}\int_0^1 \dr\mathcal{T}\, \mathcal{T}\left(z^\gamma t_{\gamma}\right)^2\int \dr^4 \rho_+\, \delta\left(1-\sum_{i=1}^4 \rho_i\right)\frac{\rho_1}{(\rho_1+\rho_2)(\rho_3+\rho_4)}\times\\
\times\exp\Big\{i\mathcal{T}z_\alpha y^\alpha+i\mathcal{T}z^\alpha\Big(-p_{1 \alpha}-(\rho_1+\rho_2)p_{2 \alpha}+(\rho_3+\rho_4)p_{3 \alpha}+(1-\rho_2)t_{\alpha}\Big)+iy^\alpha p_{1 \alpha}\\
+\frac{\rho_1\rho_3}{(\rho_1+\rho_2)(\rho_3+\rho_4)}\left(iy^\alpha t_{\alpha}-ip_{1 \alpha}t^\alpha\right)-i\left(\frac{(1-\rho_4)\rho_2}{\rho_1+\rho_2}+\rho_4\right)p_{2 \alpha}t^\alpha+i\frac{\rho_1\rho_4}{\rho_3+\rho_4}p_{1 \alpha}t^\alpha\Big\}CCC\go\, ,
\end{multline}
Here sign $\approx$ means that the dominated terms are omitted and
the following shorthand notation for the integrals is used
\begin{equation}
\int \dr^n \rho_+:=\int \dr^n\rho\, \theta(\rho_1)\ldots
\theta(\rho_n)\,.
\end{equation}

One can check that the shift symmetry constraints
\eqref{Pza}-\eqref{Pta} are fulfilled for each individual
contribution. Also one can separate $z$-dependence in the
pre-exponential as in \eqref{sep} and see that \eqref{pia} is
fulfilled as well for each individual term. Then according to the
scheme proposed in section \ref{decZ} one can decrease power of
$z$ in the pre-exponential to the linear order and by applying
$\ord_{-\infty}$ reduce the expression to the form of
\eqref{Start} with $\Pi_{\sigma}$ enjoying \eqref{PLT3}.
Therefore, the conditions of the $z$ -- dominance lemma hold
implying spin-locality of $\Upsilon^{\eta\eta}_{CCC \go}$.

%\bibliographystyle{hieeetr}

%\bibliography{Z}

%\end{document}

\end{document}